# Measuring valley polarization in two-dimensional materials with second-harmonic spectroscopy

Ho Yi Wei[1,2,3], Henrique G. Rosa[2], Ivan Verzhbitskiy[2,3], Manuel J. L. F. Rodrigues[2,4], Takashi Taniguchi[5], Kenji Watanabe[5], Goki Eda[2,3,6], Vitor M. Pereira[2,3, §], José C. Viana-Gomes[2,3,4,§]

[1]*NUS Graduate School for Integrative Sciences and Engineering, 21 Lower Kent Ridge, Singapore 119077.*
[2]*Centre for Advanced 2D Materials and Graphene Research Centre, National University of Singapore, 6 Science Drive 2, Singapore 117546.*
[3]*Department of Physics, National University of Singapore, 2 Science Drive 3, Singapore 117551.*
[4]*Center of Physics and Department of Physics, Universidade do Minho, 4710-057, Braga, Portugal.*
[5]*National Institute for Materials Science, 1-1 Namiki, Tsukuba 305-0044, Japan.*
[6]*Department of Chemistry, National University of Singapore, 3 Science Drive 3, Singapore 117543.*

A population imbalance at different valleys of an electronic system lowers its effective rotational symmetry. We introduce a technique to measure such imbalance — a valley polarization — that exploits the unique fingerprints of this symmetry reduction in the polarization-dependent second-harmonic generation (SHG). We present the principle and detection scheme in the context of hexagonal two-dimensional crystals, which include graphene-based systems and the family of transition metal dichalcogenides, and provide a direct experimental demonstration using a 2H-$MoSe_2$ monolayer at room temperature. We deliberately use the simplest possible setup, where a single pulsed laser beam simultaneously controls the valley imbalance and tracks the SHG process. We further developed a model of the transient population dynamics which analytically describes the valley-induced SHG rotation in very good agreement with the experiment. In addition to providing the first experimental demonstration of the effect, this work establishes a conceptually simple, compact and transferable way of measuring instantaneous valley polarization, with direct applicability in the nascent field of valleytronics.

In semiconductors, “valleys” refer to local extrema of the electronic band structure in momentum space, i.e. local maxima (minima) of the valence (conduction) band. Valleytronics aims to manipulate the population of charge carriers in each valley, which requires driving a population imbalance between non-equivalent valleys. The ability to induce, sustain, and control a such a “valley polarization” (VP) endows electrons with a new (valley) degree-of-freedom to encode information[1–3], beyond its charge and spin.

Systems based on graphene or transition metal dichalcogenides (TMDs) are forefront platforms to explore the promise of valleytronics[4]. Their unique electronic structure features an underlying Diracness associated with two non-equivalent valleys ($\pm K$) enabled by the hexagonal crystal structure[5], each valley carrying a finite Berry curvature[4,6]. This is compounded with their facile electrostatic (gate) tunability[7], strong and rich interaction with light[8], and the versatility offered by combining different monolayers in complex heterostructures[9]. Such features enabled the recent observation of valley Hall effect[10,11], or the optical excitation and quantum manipulation[12] of VP in this class of systems. Structures lacking inversion symmetry (e.g. monolayer TMDs or biased graphene multilayers), in particular, display an optical selection rule that allows remote control over the electronic population in each valley by shining either left ($\sigma^+$) or right ($\sigma^-$) circularly polarized light onto the sample[9,13].

Crucial to support the development of valleytronics is the ability to detect and quantify the degree of VP on a sample with spatial resolution, instantaneously, and by non-invasive and expedite means. Up to now, this has been approached using polarization-resolved photoluminescence (PL)[14–17] and the magneto-optic Kerr effect[18–20]/circular dichroism[21,22]. However, VP detection by PL requires the material to have a band gap and the readout is convoluted with the excited state lifetime[19,23]. On the other hand, for the Kerr effect to arise, the material needs to have a strong spin-orbit coupling so that a VP correlates with circular dichroism[11,19–22,24,25]. Most significantly, both shortcomings impede their use in monolayer graphene, which has neither a gap nor a practically relevant spin-orbit coupling[5].

Recent theoretical work proposed second-harmonic generation (SHG) to probe VP and valley-polarized currents in graphene and TMDs[26–30]. Governed by the second-order nonlinear susceptibility tensor $\chi^{(2)}$, SHG is a robust intrinsic feature of monolayer TMDs[31–34], but is absent in centrosymmetric crystals such as graphene or even-layered TMDs of the 2H polytype. In contrast, non-centrosymmetric crystals have finite $\chi^{(2)}$ components imposed by their point group symmetry. The point group of monolayer 2H-TMDs at equilibrium is $D_{3h}$, which implies that the non-zero components of the intrinsic $\chi^{(2)}$ are all related to a single quadratic susceptibility parameter[35,36]: $\chi_{\text{int}} \equiv \chi^{(2)}_{222} = -\chi^{(2)}_{211} = -\chi^{(2)}_{121} = -\chi^{(2)}_{112}$, where the Cartesian direction 2 is parallel to the crystal's vertical mirror symmetry. An out-of-equilibrium VP lowers this symmetry and gives rise to additional non-zero components[28,30], namely $\chi_{\text{vp}} \equiv \chi^{(2)}_{111} = -\chi^{(2)}_{122} = -\chi^{(2)}_{212} = -\chi^{(2)}_{221}$, where $\chi^{(2)}_{111} \neq \chi^{(2)}_{222}$. Consequently, the polarization-resolved SHG undergoes detectable changes according to the amount of VP induced[28,30]. For example, VP can be manifested by a chemical potential difference, $\Delta\mu$, between the valleys, and it can be shown that, to lowest order, the magnitude of $\chi_{\text{vp}}$ increases linearly with $\Delta\mu$ [26,28,30,37]. More generally, defining the degree of VP as $\Delta N \equiv N_+ - N_-$ where $N_\pm$ is the population at each of the $\pm K$ valleys, one expects $\chi_{\text{vp}} \propto \Delta N$ for a two-dimensional (constant density of states) system, whereas $\chi_{\text{int}}$ is independent of $\Delta N$.

The contributions $\chi_{\text{int}}$ and $\chi_{\text{vp}}$ to the SHG may be distinguished by polarization spectroscopy because their respective nonlinear polarizations, $\boldsymbol{P}^{(2)}$, point along orthogonal directions. Therefore, if the system has an intrinsic $\chi^{(2)}$, the emergence of VP is manifested by a rotation of the polarization-dependent SHG signal. More interestingly, in the case of graphene and other centrosymmetric crystals where $\chi_{\text{int}} = 0$, gauging VP with SHG is even simpler, because the nonlinear effect can only arise

§Corresponding authors: vpereira@nus.edu.sg, phyvjc@nus.edu.sg.

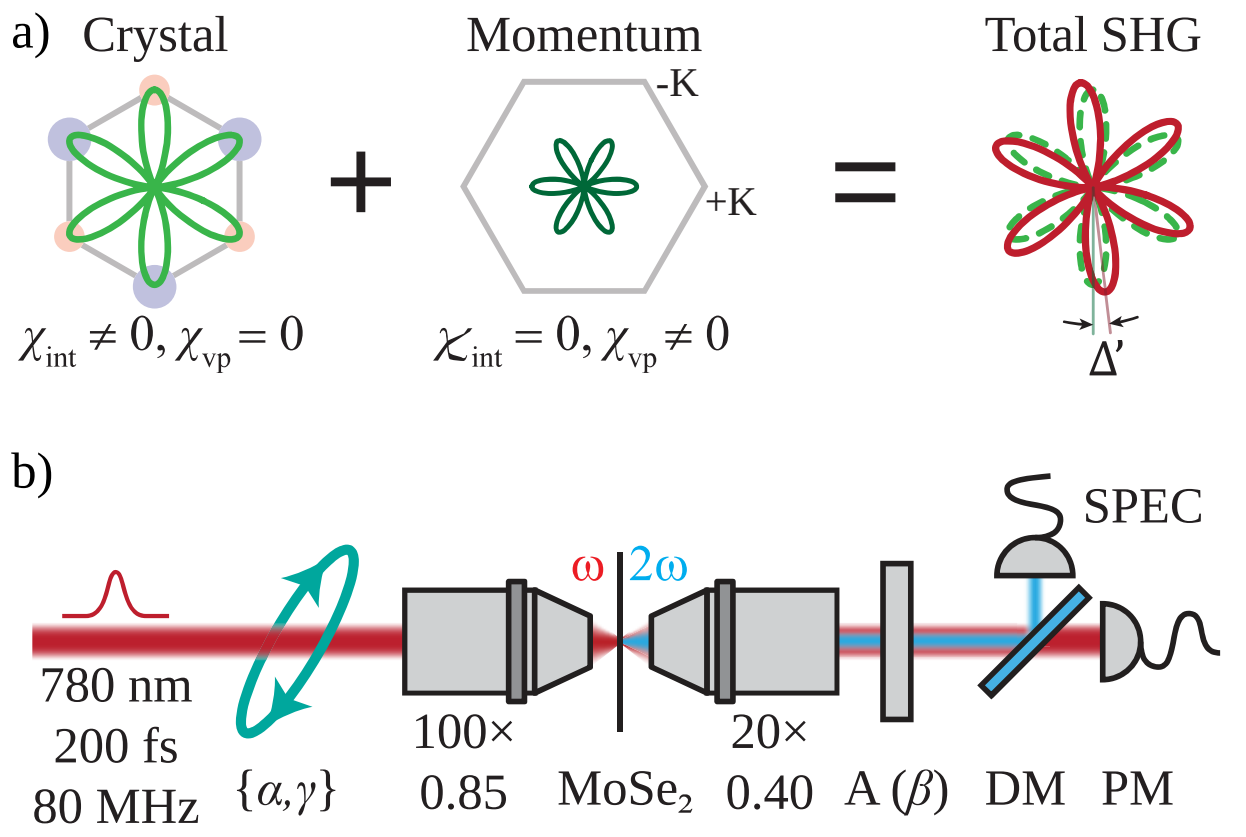

FIGURE 1: a) Illustration of the combined effect of intrinsic and VP-induced $\chi^{(2)}$ components on the total SHG pattern after the linear analyzer. A finite VP translates in a rotation ($\Delta'$) of the SHG pattern relative to that of the crystal at equilibrium. b) Experimental setup used to pump a $MoSe_2$ monolayer with an elliptically polarized pulsed laser beam (DM: dichroic mirror, PM: power meter, SPEC: spectrometer).

from $\chi_{vp}$: observation of SHG in such cases constitutes a direct detection of VP.

In this context, we report here the first experimental demonstration of SHG as an effective tool to probe VP at room temperature. In a home-built multiphoton microscope, a monolayer of molybdenum diselenide (2H-$MoSe_2$) is illuminated by a *single* laser beam that, if prepared with elliptical polarization (EP), can simultaneously induce VP and pump the SHG process. By continuously varying the beam's polarization state from linear to circular, we control the build-up of a valley imbalance which, correspondingly, causes the appearance and progressive development of a non-zero $\chi_{vp}$ contribution to $\chi^{(2)}$. This non-equilibrium component combines with the intrinsic second-order response and produces a well-defined rotation of the polarization-resolved SHG pattern[34,38,39]. The rotation directly yields the magnitude of $\chi_{vp}$. As the same laser beam is used to both pump and probe (via SHG) the valley imbalance, this technique makes use of a simple setup that bypasses the need of superimposing two separate beams either in time or space. It may thus be used to implement very compact, fast and robust protocols to track the valley degree of freedom.

**Linearly polarized SHG** – To best appreciate the nature of the effect and the underlying principle of operation, consider first the case where a valley-polarized $MoSe_2$ crystal is excited with *linearly* polarized light. With the electric field vector on the crystal plane, the nonlinear polarization vector is given by[28,30]

$$\begin{bmatrix} P^{(2)}_{ZZ} \\ P^{(2)}_{AC} \end{bmatrix} \propto \begin{bmatrix} \chi_{vp} & -\chi_{vp} & -\chi_{int} \\ -\chi_{int} & \chi_{int} & -\chi_{vp} \end{bmatrix} \begin{bmatrix} E^2_{ZZ} \\ E^2_{AC} \\ 2E_{ZZ}E_{AC} \end{bmatrix}, \quad (1)$$

where $E_{ZZ}$ and $E_{AC}$ are the components of the incoming field along the zig-zag (ZZ) and armchair (AC) directions, respectively. The existence of a non-equilibrium VP is reflected in the presence of the $\chi_{vp}$ terms that are otherwise absent in the intrinsic $\chi^{(2)}$ tensor at equilibrium. A simple inspection of Equation (1) shows that the intensity of the SHG signal is independent of the incoming polarization angle, $\beta$, measured with respect to the AC direction. However, if the SHG signal is analyzed with a polarizer aligned in the same direction $\beta$ as the pump field, the intensity becomes $S(\beta) \propto |\chi_{int}\cos 3\beta - \chi_{vp}\sin 3\beta|^2$, which explicitly depends on both the intrinsic and VP-induced contributions to $\chi^{(2)}$. One easily sees that the pattern described by $S(\beta)$ in a polar representation as a function of $\beta$ corresponds to a scaling and rotation of the six-fold pattern we obtain with $\chi_{vp} = 0$ (see Figure 1a). The rotation angle is $\Delta' \approx [\chi_{vp}/(3\chi_{int})]\cos\phi$, where $\phi$ is the relative phase between the complex $\chi_{int}$ and $\chi_{vp}$ (near a resonance, which is our case of interest, their phases are expected to be similar; thus $\phi \approx 0$ and we consider $\{\chi_{vp}, \chi_{int}\} \in \mathbb{R}$ henceforth).

The readout method just discussed is illustrated in Figure 1a; it provides an intuitive way of measuring VP by tracking the rotation angle $\Delta'$. The signal is characterized by a six-petal pattern and, other than the rotation and scaling induced by a finite VP, is analogous to the most commonly used SHG technique to find the lattice orientation of these crystals.[31,34,40,41]; but, with a finite VP, the amount of rotation undergone by the flower-shaped pattern is directly related to the degree of VP. A simple variation of this protocol increases the rotation (hence the sensitivity) by a factor of three: Holding the incoming linear polarization along a fixed direction and rotating the analyzer by the *relative* angle $\beta$, the recorded SHG signal reads $S(\beta) \propto |\chi_{int}\cos\beta - \chi_{vp}\sin\beta|^2$. This yields a two-fold pattern as a function of $\beta$ (cf. Figure 2b) which rotates by the angle

$$\Delta = \tan^{-1}\left(\frac{\chi_{vp}}{\chi_{int}}\right) \approx \frac{\chi_{vp}}{\chi_{int}} \quad (2)$$

with respect to its position in the absence of VP (the last result is valid if $|\chi_{vp}| \ll |\chi_{int}|$).

We thus see that the presence of a VP can be optically detected and quantified in a rather simple SHG setup. By allowing that in a non-invasive and remote way, this approach can have many applications to directly monitor the valley degree of freedom in a number of valleytronics applications. While the previous discussion assumes that a VP has been established in the system by independent external means, we now show that the same simple setup can be used to probe the VP response of a crystal in equilibrium.

**Simultaneous pumping and probing VP** – As we wish to demonstrate the rotation effect described above in a standalone crystal in equilibrium, we need a controllable way of inducing a population imbalance in the two valleys $\pm K$. This can be achieved by preparing the incoming beam in a state of elliptical polarization (EP), which we characterize by the helicity angle : $\sin 2\gamma = (I_+ - I_-)/(I_+ + I_-)$, where $I_\pm$ are the intensities of the two circular components in the beam. In a generic state with helicity $0 < |\gamma| < 45°$ the two intensities $I_\pm$ are different; since in $MoSe_2$ the $\sigma^+$ ($\sigma^-$) photons are selectively absorbed by particle-hole excitations only at the $+K$ ($-K$) valley[14–17], EP light naturally induces different exciton populations at each valley by an amount that is controlled by the value of $\gamma$. At the same time, as long as $\gamma \neq \pm 45°$, an EP beam still drives SHG with the symmetries discussed above and, therefore, the rotation of the nonlinear signal can be tracked to monitor the VP in the system. This allows one to simultaneously pump a VP on $MoSe_2$ and measure the magnitude of that VP, with the same laser beam.

A pulsed laser beam (Toptica FemtoFiber pro NIR, $\lambda =$ 780 nm, $\hbar\omega = 1.59$ eV, $\tau = 200$ fs full-width at half-maximum pulse duration, 80 MHz repetition rate) was directed at a $MoSe_2$ monolayer according to the optical setup illustrated in Figure 1b. The photon energy is quasi-resonant with the lowest-lying exciton state (~19 meV above the PL peak at $E_A = 1.571$ eV, see Figure S1c in the Supplementary Information). The monolayer was obtained by mechanical exfoliation from a 2H-$MoSe_2$ single-crystal, transferred to a fused silica substrate and encapsulated by 20 nm of hexagonal boron nitride to minimize environmental degradation. In addition to SHG spectroscopy, the single-layer character was validated by the Raman and PL spectroscopies (Supplementary Section 1). All our optical measurements were performed under ambient conditions at room temperature. The incoming polarization was manipulated by a set of polarizers and waveplates. The semi-major axis of the polarization ellipse is oriented along $\alpha$ (see Figure 2a), which is kept constant and aligned to the AC direction of the sample, while $\gamma$ is varied from $-30°$ to $+30°$. For each $\gamma$, the analyzer is rotated 360° to analyze the SHG signal before reaching a spectrometer. At the sample, the pump laser had an average power of few mW, focused by a 100× objective down to spot diameter of 1 µm.

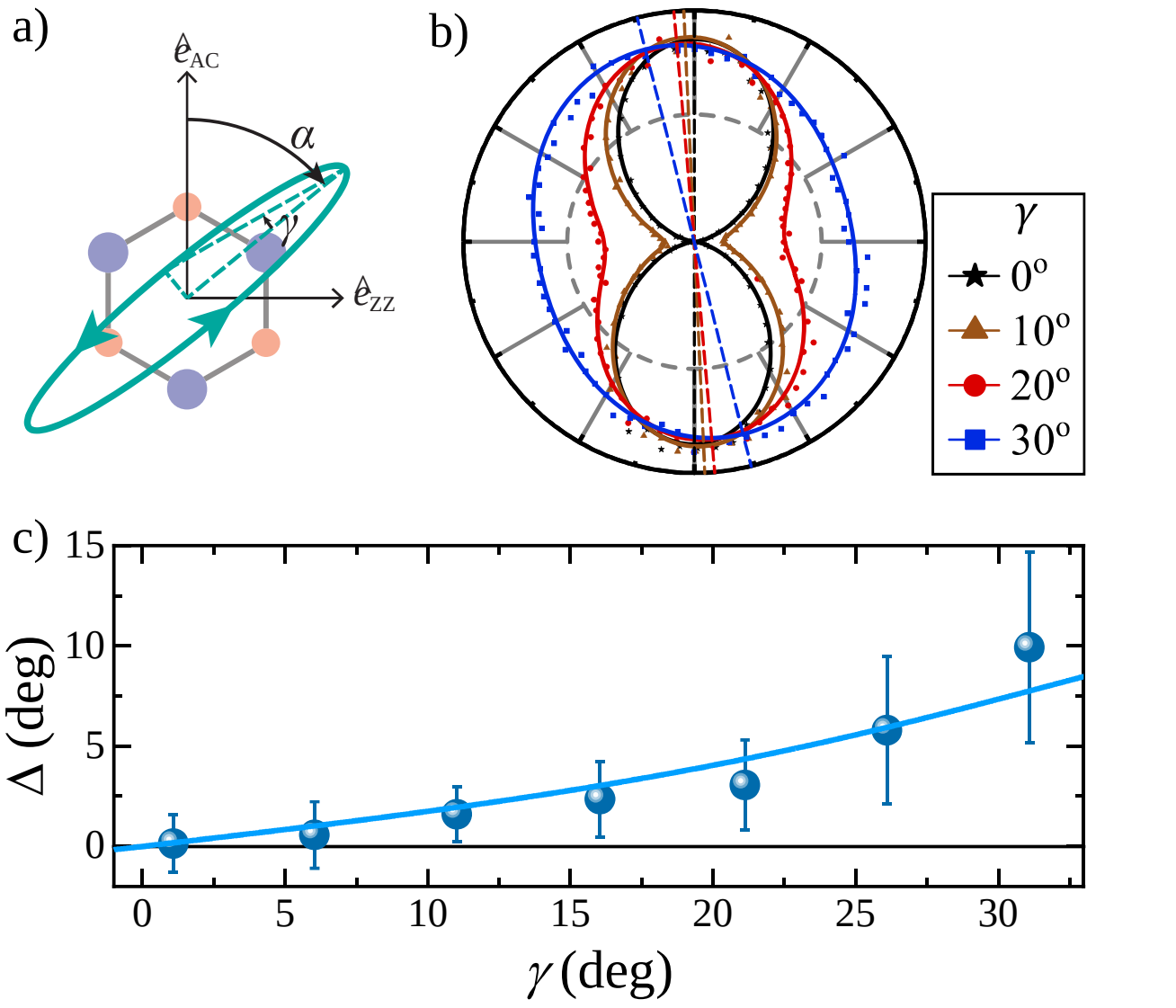

FIGURE 2: a) Parameters $\alpha$ and $\gamma$ that define the elliptical polarization state of the incoming light beam, and the orientation of the semi-major axis relative to the crystal lattice. b) Representative measurements of the (normalized) SHG signal as a function of analyzer angle (the polar angle of the plot) for different helicities $\gamma$. Solid lines are the best fit to the experimental points. The dashed straight lines highlight the progressive rotation of the twofold pattern with increasing $\gamma$. c) The rotation $\Delta$ of the SHG pattern as a function of helicity. In (b) and (c) the peak irradiance at the sample was 2.1 GW cm$^{-2}$.

A set of data for different values of $\gamma$ is shown in the polar plots of Figure 2b. As expected for linearly polarized light ($\gamma = 0$), the analyzed SHG signal displays the pattern $S(\beta) \propto |\chi_{\mathrm{int}}|^2 \cos^2(\beta)$. A finite $\gamma$ introduces two effects, which are evident in this figure: the minimum of the SHG signal is no longer at zero, and the two-fold pattern undergoes a finite rotation that becomes more pronounced with increasing $\gamma$. While the first effect is a simple consequence of exciting the SHG with EP light[28,30], the rotation is a fingerprint of VP due to the different rate at which $\sigma^+$ and $\sigma^-$ photons are absorbed by the sample. Equation (2) shows that the amount of rotation $\Delta$ relative to the pattern for $\gamma = 0$ provides direct access to the non-equilibrium contribution ($\chi_{\mathrm{vp}}$) to $\chi^{(2)}$. The magnitude of $\chi_{\mathrm{vp}}$, in turn, provides a direct measure of the valley population imbalance.

A summary of the SHG rotation as a function of $\gamma$ is shown in Figure 2c. At small $\gamma$, while the incoming polarization is still approximately linear, the population imbalance is small because $I_+ - I_- \ll I_+ + I_-$. Since, as pointed out earlier, in this regime $\chi_{\mathrm{vp}} \propto \Delta N$, one expects to see $\chi_{\mathrm{vp}} \propto I_+ - I_- \propto \sin 2\gamma \approx 2\gamma$ and, according to Equation (2), one anticipates $\Delta \propto \gamma$. The data reflects precisely this behavior and, as a result, confirms the predicted linear relation $\chi_{\mathrm{vp}} \propto \Delta N$ at moderate valley pumping. To understand the full response curve at high $\gamma$ in this single-beam setup we must consider the transient processes and partial saturation that takes place within each laser pulse, which we address below. Note that the error bars in Figure 2c grow with increasing $\gamma$ because the SHG pattern becomes progressively more isotropic (cf. Figure 2b), which increases the error in the numerical fitting (Supplementary Section 2) that is required to extract $\Delta$.

Overall, these data establish two things: (i) that SHG is indeed an effective means of characterizing the degree of VP in the system, and (ii) that a single-beam setup can be used, not only to probe, but also to simultaneously drive VP in a crystal from equilibrium. The latter aspect allows, for example, to characterize materials according to the valley population imbalance, $\Delta N$, that is produced by a given helicity $\gamma$; this response defines a material's "valley susceptibility", $k_v$, according to $\Delta N = k_v\, \gamma$.

**Transient and effective response** – Resolving the SHG pattern rotation with sufficient signal-to-noise ratio requires high irradiance, while, on the other hand, to efficiently pump VP (which is a linear optical absorption process), the absorption must not be saturated. As our experiment uses the same source to drive those two (1$^{\mathrm{st}}$ and 2$^{\mathrm{nd}}$ order) processes, it would seem to require a delicate compromise. The situation is facilitated by the use of an ultrafast pulsed laser beam (for which the recorded SHG signal is integrated over many pulses), and the fact that valley populations undergo complex transients within each pulse's duration ($\tau$) due to various decay pathways on time-scales comparable with $\tau$ [42]. To better understand these dynamics and how it defines the total SHG signal, we developed a model that considers the temporal evolution of the valley imbalance within one pulse and allows us to describe the SHG rotation in terms of the helicity $\gamma$ at arbitrary power.

The quasi-resonance of the incoming photons with the A exciton, combined with the large exciton binding energy[43], justifies modeling the key processes with a pair of two-level systems $\{|0\rangle, |\pm\rangle\}$, where $|0\rangle$ represents the ground state on either valley, and $|\pm\rangle$ an A exciton belonging to the $\pm K$ valley. We consider the population rate equations,

$$\begin{aligned} \frac{\mathrm{d}N_\pm}{\mathrm{d}t} = \; & -(\Gamma + \Gamma_{\mathrm{nr}})N_\pm - \frac{\sigma(\omega)}{n\,\hbar\omega} I_\pm \big(N_\pm - N_{0,\pm}\big) \\ & -A N_\pm^2 \pm \Gamma_v (N_\mp - N_\pm), \end{aligned} \qquad (3)$$

where $N_i$ represents the population in state $|i\rangle$, $\Gamma_{(\mathrm{nr})}$ is the (non) radiative recombination rate, $AN_\pm$ the exciton-exciton annihilation rate[44], $\Gamma_v$ the intervalley recombination rate, $I_\pm$ is the intensity of each circular component of the incoming light, $\sigma(\omega)$ is the $MoSe_2$ absorption cross-section, $n$ its refractive index, $\hbar\omega$ the photon energy, and $N_\pm + N_{0,\pm} = N_T$ is the total density of available particle-hole excitations, which is an intrinsic property of $MoSe_2$ (details in Supplementary Section 5) The laser pulse is modeled as $I(t) = I_0 f(t)$, where $f(t)$ is a Gaussian function characterized by the experimental width, $\tau$. The last term in Equation (3) describes intervalley exciton transitions which, in the specific case of $MoSe_2$, are known to contribute negligibly to the total exciton linewidth ($\Gamma_v \ll \Gamma + \Gamma_{\mathrm{nr}}$), even at room temperature[42]. Moreover, in our experiment the photon detuning is only $h\nu - E_A \simeq 19$ meV, below the 38 meV necessary to activate phonon-assisted intervalley relaxation[17]. Therefore, we set $\Gamma_v = 0$ (justified within the 200 fs width of the laser pulse) and solved the two Equations (3) with the following parameters applicable to $MoSe_2$[42,44,45]: $\Gamma$ = 3.3 ps$^{-1}$, $\Gamma_{\mathrm{nr}}$ = 26.1 ps$^{-1}$, $A =$ 0.33 cm$^2$ s$^{-1}$, $\sigma(\omega) = N_T^{-1}(1 - \ln 0.06)$ where 0.06 is the absorption coefficient at 780 nm, and $N_T$ = 2.1×10$^{12}$ cm$^{-2}$ (estimated from the laser linewidth and the electronic density of states; see Supplementary Section 5.1).

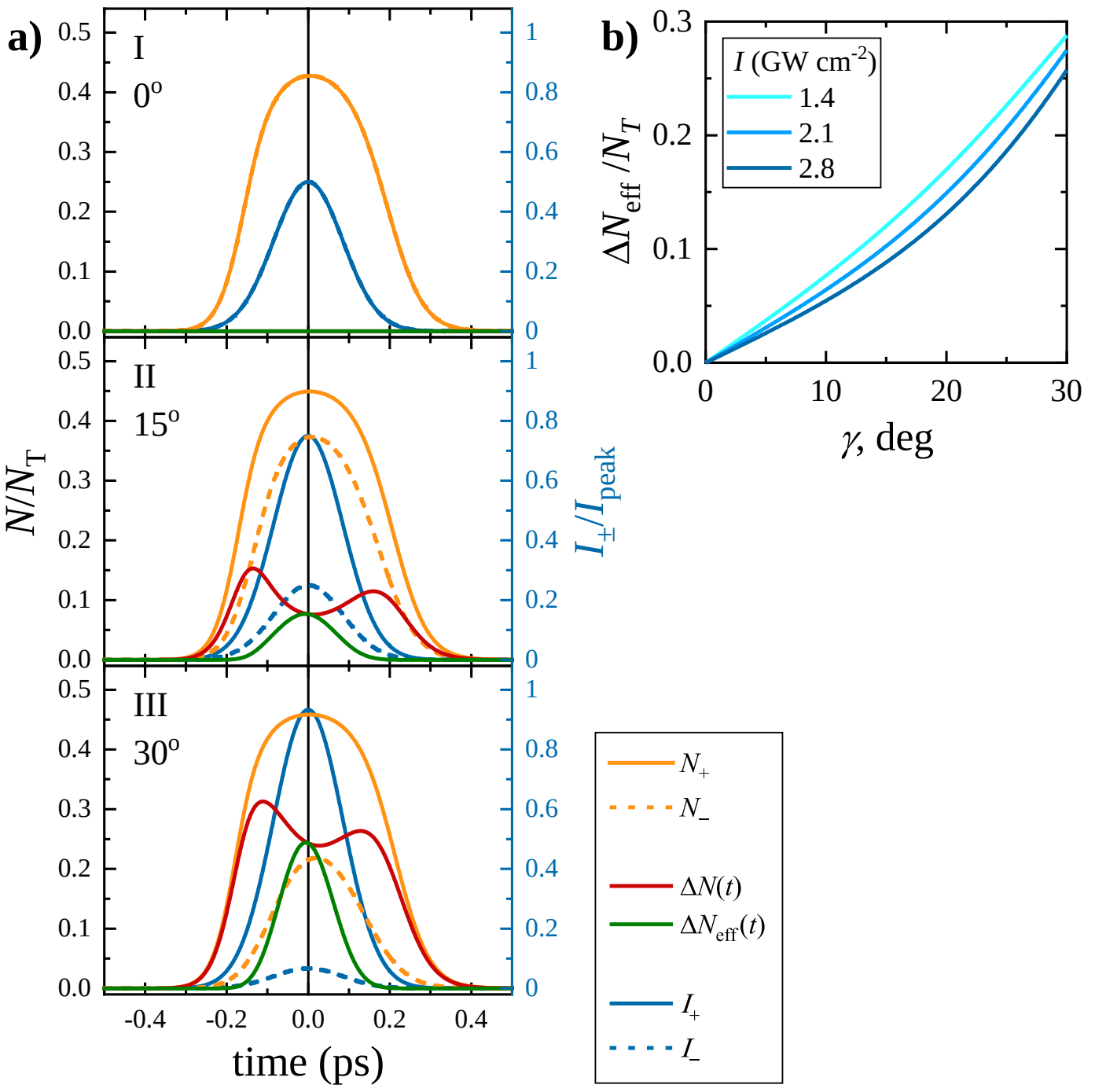

FIGURE 3: a) Computed exciton populations at the $+K$ ($N_+$, solid orange curve) and $-K$ ($N_-$, dashed orange curve) valleys for $I_0$ = 2.8 GW cm$^{-2}$ and $\gamma = 0°$ (I), 15° (II), and 30° (III). $I_+$ (solid blue curve) and $I_-$ (dashed blue curve) are the $\sigma^+$ and $\sigma^-$ intensity components of the laser pulse, respectively. The instantaneous population imbalance $\Delta N(t)$ is plotted in red while $\Delta N_{eff}(t) \equiv \Delta N(t) f^2(t)$ is shown in green. b) Pulse-averaged population imbalance obtained with Equation (4) as a function of $\gamma$, and for a range of intensities covering those used in the experiments. It is an odd function of $\gamma$.

Figure 3a shows the computed transient populations for different helicities $\gamma$ of the incoming photons, as well as the imbalance $\Delta N(t) \equiv N_+(t) - N_-(t)$ (details in Supplementary Section 5.2). [Note that setting $\Gamma_v = 0$ decouples the valleys and $\Delta N(t)$ is controlled only by $I_+(t) - I_-(t)$, which are also shown in the figure.] Even though these results were obtained with a peak pulse intensity ($I_0$) higher than the CW saturation ($I_{sat}$) by the factor of $I_0/I_{sat} \sim 10$ (see Supplementary Figure S10) no saturation of the transient populations occurs since $N_\pm(t) < N_T/2$ at all times. With increasing helicity a finite $\Delta N(t)$ obtains throughout the whole pulse (red trace in panels II and III, Figure 3a). Its double-peak structure with a dip at the center arises because, with increasing $\gamma$, one of the populations ($N_+$) approaches the saturation limit ($N_\pm = N_T/2$) before the pulse reaches peak intensity; on that nonlinear regime, $N_+$ increases at a slower rate than $N_-$, which results in the central dip in $\Delta N(t)$. The key observation is that $\Delta N(t)$ remains *finite* throughout the entire pulse, even for the large irradiances required for SHG (see Figure S6c in SI). It is thus appropriate to introduce a pulse-averaged population imbalance,

$$\Delta N_{eff} \equiv \frac{\int_{-\infty}^{\infty} \Delta N(t) f^2(t) dt}{\int_{-\infty}^{\infty} f^2(t) dt}, \tag{4}$$

which defines the effective population imbalance that determines the amount of rotation in the SHG as recorded at the detector. In addition to deriving the above result, in Supplementary Section 5.2 we show that $\Delta N_{eff}$ depends on the beam power and helicity according to

$$\frac{\Delta N_{eff}}{N_T} = \frac{b I_0/I_m \sin 2\gamma}{(b I_0/I_m + 1)^2 - (b I_0/I_m \sin 2\gamma)^2} \tag{5}$$

where $I_m = n\hbar\omega(\Gamma + \Gamma_{nr})/\sigma(\omega)$ = 7.6 MW cm$^{-2}$ and $b$ = 0.0122. With $b = 1$, this expression reduces to the steady state solution of Equations (3) (Supplementary Section 5.1). Despite the rich transient seen in Figure 3a, Equation (5) shows that the dependence of the valley imbalance with laser power and helicity can still be captured by the steady state functional dependence, but with a rescaled peak power: $I_0 \to b I_0$ (note that $b$ is a parameter independent of $I_0$ and $\gamma$; see Supplementary Section 5.2).

Expression (5) is plotted in Figure 3b and is seen to reproduce the experimental trend of Figure 2c. To make this more explicit and quantitative we recall that, since $\chi_{vp} \propto \Delta N_{eff}$, we can write $\chi_{vp}/\chi_{int} = \kappa\, \Delta N_{eff}/N_T$ and combine Equations (2) and (5) to obtain the variation of $\Delta$ with $\gamma$ (the constant $\kappa$ can in principle be determined from a microscopic model of the optical response). Figure 4 shows the outcome of applying this procedure to fit the experimental rotation $\Delta(\gamma)$ at different laser power, with the constant $\kappa$ as fitting parameter. We can see that this description captures the experimental behavior very well in Figure 4. The fit in Figure 4 yields $\kappa = 0.47 \pm 0.02$. This value means that if, for example, one induces a VP in the system of $\Delta N_{eff}/N_T = 10\%$, the valley-induced contribution to $\chi^{(2)}$ is expected to be $\chi_{vp}/\chi_{int} \approx 4.7\%$.

**Discussion and conclusions** – Since SHG requires large irradiances ($I_0$), one could think that the observation of SHG rotation might be incompatible with efficient valley pumping and, thus, with a single-beam setup (for example, if $I_0$ is so large that absorption is saturated, the valley imbalance disappears). On the other hand, the results above show that a pulsed source allows a comfortable range of intensities where that compromise can be fulfilled. This is due to the ultrafast transient of the laser pulse and valley populations, which pushes the onset of saturation to values of $I_0$ much above $I_{sat}$ under CW conditions[45] (see Supplementary Figure S10). We elaborate this in detail in Supplementary Section 5.2 and Supplementary Figures S7 to S9. We show, in particular, the existence of an optimal intensity, $I_{opt}$, which provides the highest effective VP (highest $\Delta N_{eff}$) at a given temperature ($I_{opt} \sim 1.3$ GW cm$^{-2}$ at 300 K, Supplementary Figure S9). In terms of the plots shown in Figure 3, the value of $I_{opt}$ corresponds to the threshold beyond which the central dip appears in the traces of $\Delta N(t)$; increasing $I_0$ past $I_{opt}$ brings the magnitude of $\Delta N_{eff}$ down as the dip intensifies. For this reason, the experimental intensities were chosen near the predicted $I_{opt}$ for our material parameters (cf. Figure 4). Also important is, of course, the strong coupling of light to $MoSe_2$ (and 2D materials in general), which ensures a detectable nonlinear signal well before the onset of saturation.

It is important to stress, however, that this compromise is *specific* to using the *same* laser source for *both* optical processes, which is an extreme application of the principle of using SHG to quantify VP — this work establishes its applicability to such an extreme setting as well. In practice, no such constraint exists in all other scenarios where VP arises by other means (valley Hall effect, valley-polarized carrier injection, pumping from another laser, etc.): In such cases, we need only to track the SHG pattern; moreover, that can be done entirely with linearly polarized light, according to the much simpler protocol described earlier under "Linearly polarized SHG".

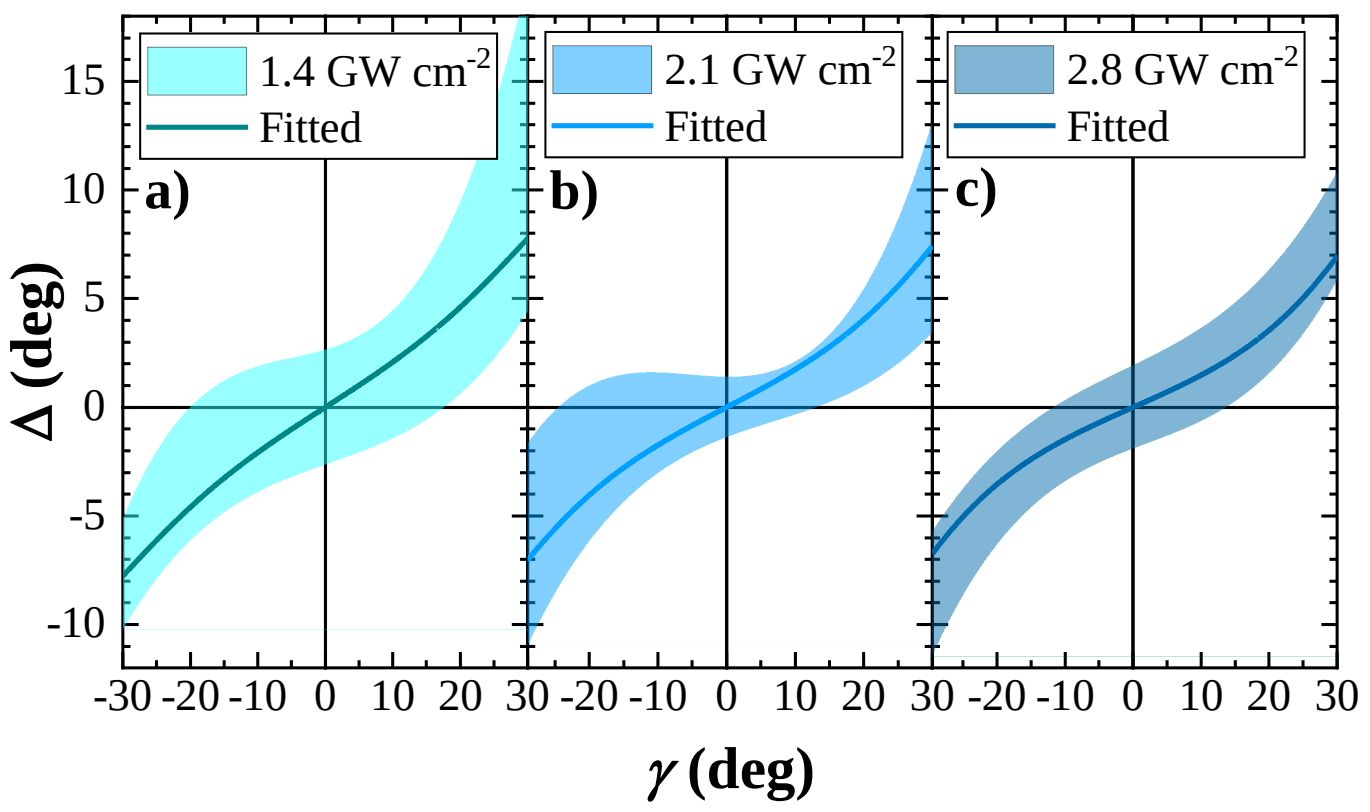

FIGURE 4: SHG rotation angle ($\Delta$) against $\gamma$ for different laser intensities. The shaded regions reflect the span of the experimental error bars (data point in the middle for each $\gamma$), as in Figure 2c. As discussed in the text, larger $\Delta$ indicates stronger VP. The solid lines are the best fit of each dataset to Equations (2) and (5).

One final remark relates to the Mott transition between the “exciton gas” and “electron-hole liquid” phases[46], which occurs when the density of particle-hole pairs exceeds $\sim 1/a_X^2$ ($a_X$ is the exciton radius), and which has been recently established theoretically and experimentally across the family of 2H TMDs[47–52]. At room temperature, the Mott transition is predicted[47–49] to occur at $\sim 10^{13}\ \mathrm{cm}^{-2}$ in $MoS_2$, possibly higher in $MoSe_2$ on account of its larger binding energy[43] and consistent with recent measurements in $MoSe_2/WSe_2$ heterobilayers[53]. Although the peak power ~1–3 $\mathrm{GW/cm^2}$ used in our experiment promotes high exciton densities, we seem to remain below the transition to the liquid phase given the agreement in Figure 4 between the observed and predicted behavior of the SHG rotation. (This is possibly aided by the small, but not insignificant, detuning from the excitonic resonance: $h\nu - E_A = 19$ meV while the pulse HWHM is $\approx 10$ meV) At any rate, we stress that the fingerprint of nonzero VP in the SHG should remain even above the Mott threshold, simply because a valley imbalance always breaks the crystal’s mirror symmetry, thereby inducing a finite $\chi_{\mathrm{vp}}$, irrespective of the nature of the excited populations. In fact, under asymmetric valley population, we expect correspondingly asymmetric redshifts of the exciton levels and bandgap near the Mott transition[47,48]; this should greatly reduce intervalley relaxation for lack of phase space, thus sustaining VP for longer times and/or larger laser detuning.

In conclusion, we demonstrated that SHG spectroscopy can be a powerful tool for valleytronics, allowing to track and quantify the degree of VP. A single-beam SHG measurement allows simple protocols for remote characterization, including the ability to spatially map/scan non-uniform valley accumulation or diffusion, by analogy with the use of Kerr rotation spectroscopy in spintronics[24]. It could be especially suited to characterize the transport of valley currents in graphene, where recent experiments demonstrate the ability to inject spin-valley-polarized currents from adjacent TMDs at room temperature[54,55]. Moreover, as a parametric process, SHG is sensitive to the instantaneous degree of VP which, given the fast valley relaxation processes[12], is an important advantage for operation at room temperature and/or photon energies with large detuning from the exciton levels, where other techniques such as PL stand at disadvantage[19].

**Acknowledgements** – H.Y.W. acknowledges the support from NUS Graduate School for Integrative Sciences and Engineering. G.E. acknowledges financial support from the National Research Foundation of Singapore (NRF Research Fellowship NRF-NRFF2011-02 and medium-sized centre programme) and Ministry of Education of Singapore (AcRF Tier 2 MOE2015-T2-2-123). V.M.P. acknowledges financial support from the Ministry of Education of Singapore (FRC AcRF Tier 1 R-144-000-386-114). J.C.V.G. acknowledges financial support from CA2DM through National Research Foundation of Singapore (NRF-CRP Grant No. R-144-000-295-281).

## References

[1] O. Gunawan, Y.P. Shkolnikov, K. Vakili, T. Gokmen, E.P. De Poortere, and M. Shayegan, Phys. Rev. Lett. **97**, 186404 (2006).

[2] Y.S. Ang, S.A. Yang, C. Zhang, Z. Ma, and L.K. Ang, Phys. Rev. B **96**, 245410 (2017).

[3] A. Rycerz, J. Tworzydło, and C.W.J.J. Beenakker, Nat. Phys. **3**, 172 (2007).

[4] J.R. Schaibley, H. Yu, G. Clark, P. Rivera, J.S. Ross, K.L. Seyler, W. Yao, and X. Xu, Nat. Rev. Mater. **1**, (2016).

[5] K.S. Novoselov, A.K. Geim, F. Guinea, N.M.R. Peres, and A.H. Castro Neto, Rev. Mod. Phys. **81**, 109 (2009).

[6] H. Yu, X. Cui, X. Xu, and W. Yao, Natl. Sci. Rev. **2**, 57 (2015).

[7] S. Manzeli, D. Ovchinnikov, D. Pasquier, O. V. Yazyev, and A. Kis, Nat. Rev. Mater. **2**, (2017).

[8] K.F. Mak and J. Shan, Nat. Photonics **10**, 216 (2016).

[9] X. Xu, W. Yao, D. Xiao, and T.F. Heinz, Nat. Phys. **10**, 343 (2014).

[10] K.F. Mak, K.L. McGill, J. Park, and P.L. McEuen, Science. **344**, 1489 (2014).

[11] J. Lee, K.F. Mak, and J. Shan, Nat. Nanotechnol. **11**, 421 (2016).

[12] K.F. Mak, D. Xiao, and J. Shan, Nat. Photonics **12**, 451 (2018).

[13] W. Yao, D. Xiao, and Q. Niu, Phys. Rev. B **77**, 235406 (2008).

[14] T. Cao, G. Wang, W. Han, H. Ye, C. Zhu, J. Shi, Q. Niu, P. Tan, E. Wang, B. Liu, and J. Feng, Nat. Commun. **3**, 887 (2012).

[15] H. Zeng, J. Dai, W. Yao, D. Xiao, and X. Cui, Nat. Nanotechnol. **7**, 490 (2012).

[16] K.F. Mak, K. He, J. Shan, and T.F. Heinz, Nat. Nanotechnol. **7**, 494 (2012).

[17] G. Kioseoglou, A.T. Hanbicki, M. Currie, A.L. Friedman, and B.T. Jonker, Sci. Rep. **6**, 25041 (2016).

[18] C.R. Zhu, K. Zhang, M. Glazov, B. Urbaszek, T. Amand, Z.W. Ji, B.L. Liu, and X. Marie, Phys. Rev. B **90**, 161302 (2014).

[19] W.-T. Hsu, Y.-L. Chen, C.-H. Chen, P.-S. Liu, T.-H. Hou, L.-J. Li, and W.-H. Chang, Nat. Commun. **6**, 8963 (2015).

[20] T. Yan, S. Yang, D. Li, and X. Cui, Phys. Rev. B **95**, 241406 (2017).

[21] Q. Wang, S. Ge, X. Li, J. Qiu, Y. Ji, J. Feng, and D. Sun, ACS Nano **7**, 11087 (2013).

[22] C. Mai, A. Barrette, Y. Yu, Y.G. Semenov, K.W. Kim, L. Cao, and K. Gundogdu, Nano Lett. **14**, 202 (2014).

[23] G. Wang, E. Palleau, T. Amand, S. Tongay, X. Marie, and B. Urbaszek, Appl. Phys. Lett. **106**, 112101 (2015).

[24] Y.K. Kato, R.C. Myers, A.C. Gossard, and D.D. Awschalom, Science. **306**, 1910 (2004).

[25] C. Jin, E.C. Regan, D. Wang, M. Iqbal Bakti Utama, C. Yang, J. Cain, Y. Qin, Y. Shen, Z. Zheng, K. Watanabe, T. Taniguchi, S. Tongay, A. Zettl, and F. Wang, Nat. Phys. (2019).

[26] T.O. Wehling, A. Huber, A.I. Lichtenstein, and M.I. Katsnelson,

Phys. Rev. B **91**, 041404 (2015).

[27] J. Cheng, D. Huang, T. Jiang, Y. Shan, Y. Li, S. Wu, and W.-T. Liu, Opt. Lett. **44**, 2141 (2019).

[28] F. Hipolito and V.M. Pereira, 2D Mater. **4**, 021027 (2017).

[29] Y.K. Luo, J. Xu, T. Zhu, G. Wu, E.J. McCormick, W. Zhan, M.R. Neupane, and R.K. Kawakami, Nano Lett. **17**, 3877 (2017).

[30] F. Hipolito and V.M. Pereira, 2D Mater. **4**, 039501 (2017).

[31] Y. Li, Y. Rao, K.F. Mak, Y. You, S. Wang, C.R. Dean, and T.F. Heinz, Nano Lett. **13**, 3329 (2013).

[32] N. Kumar, S. Najmaei, Q. Cui, F. Ceballos, P.M. Ajayan, J. Lou, and H. Zhao, Phys. Rev. B **87**, 161403 (2013).

[33] H.G. Rosa, L. Junpeng, L.C. Gomes, M.J.L.F. Rodrigues, S.C. Haur, and J.C. V Gomes, Adv. Opt. Mater. **6**, 1701327 (2018).

[34] H.G. Rosa, Y.W. Ho, I. Verzhbitskiy, M.J.F.L. Rodrigues, T. Taniguchi, K. Watanabe, G. Eda, V.M. Pereira, and J.C. V Gomes, Sci. Rep. **8**, 10035 (2018).

[35] Y.R. Shen, *The Principles of Nonlinear Optics* (Wiley, Chichester;New York;, 2003).

[36] R.W. Boyd, *Nonlinear Optics*, 3rd ed. (Academic Press, Burlington, MA, 2008).

[37] L.E. Golub and S.A. Tarasenko, Phys. Rev. B **90**, 201402 (2014).

[38] A. Autere, H. Jussila, Y. Dai, Y. Wang, H. Lipsanen, and Z. Sun, Adv. Mater. **30**, 1705963 (2018).

[39] A. Autere, H. Jussila, A. Marini, J.R.M. Saavedra, Y. Dai, A. Säynätjoki, L. Karvonen, H. Yang, B. Amirsolaimani, R.A. Norwood, N. Peyghambarian, H. Lipsanen, K. Kieu, F.J.G. de Abajo, and Z. Sun, Phys. Rev. B **98**, 115426 (2018).

[40] L.M. Malard, T. V. Alencar, A.P.M. Barboza, K.F. Mak, and A.M. de Paula, Phys. Rev. B **87**, 201401 (2013).

[41] J. Ribeiro-Soares, C. Janisch, Z. Liu, A.L. Elías, M.S. Dresselhaus, M. Terrones, L.G. Cançado, and A. Jorio, 2D Mater. **2**, 045015 (2015).

[42] M. Selig, G. Berghäuser, A. Raja, P. Nagler, C. Schüller, T.F. Heinz, T. Korn, A. Chernikov, E. Malic, and A. Knorr, Nat. Commun. **7**, 13279 (2016).

[43] G. Wang, A. Chernikov, M.M. Glazov, T.F. Heinz, X. Marie, T. Amand, and B. Urbaszek, Rev. Mod. Phys. **90**, 21001 (2018).

[44] N. Kumar, Q. Cui, F. Ceballos, D. He, Y. Wang, and H. Zhao, Phys. Rev. B **89**, 125427 (2014).

[45] Z. Nie, C. Trovatello, E.A.A. Pogna, S. Dal Conte, P.B. Miranda, E. Kelleher, C. Zhu, I.C.E. Turcu, Y. Xu, K. Liu, G. Cerullo, and F. Wang, Appl. Phys. Lett. **112**, 031108 (2018).

[46] J. Shah, M. Combescot, and A.H. Dayem, Phys. Rev. Lett. **38**, 1497 (1977).

[47] A. Steinhoff, M. Rösner, F. Jahnke, T.O. Wehling, and C. Gies, Nano Lett. **14**, 3743 (2014).

[48] L. Meckbach, T. Stroucken, and S.W. Koch, Appl. Phys. Lett. **112**, (2018).

[49] A. Rustagi and A.F. Kemper, Nano Lett. **18**, 455 (2018).

[50] A. Chernikov, C. Ruppert, H.M. Hill, A.F. Rigosi, and T.F. Heinz, Nat. Photonics **9**, 466 (2015).

[51] Y. Yu, A.W. Bataller, R. Younts, Y. Yu, G. Li, A.A. Puretzky, D.B. Geohegan, K. Gundogdu, and L. Cao, ACS Nano **13**, 10351 (2019).

[52] A.W. Bataller, R.A. Younts, A. Rustagi, Y. Yu, H. Ardekani, A. Kemper, L. Cao, and K. Gundogdu, Nano Lett. **19**, 1104 (2019).

[53] J. Wang, J. Ardelean, Y. Bai, A. Steinhoff, M. Florian, F. Jahnke, X. Xu, M. Kira, J. Hone, and X.-Y. Zhu, Sci. Adv. **5**, eaax0145 (2019).

[54] Y.K. Luo, J. Xu, T. Zhu, G. Wu, E.J. McCormick, W. Zhan, M.R. Neupane, and R.K. Kawakami, Nano Lett. **17**, 3877 (2017).

[55] C.K. Safeer, J. Ingla-Aynés, F. Herling, J.H. Garcia, M. Vila, N. Ontoso, M.R. Calvo, S. Roche, L.E. Hueso, and F. Casanova, Nano Lett. **19**, 1074 (2019).

Supplementary Information for

# Measuring valley polarization in two-dimensional materials with second-harmonic spectroscopy

Ho Yi Wei[1,2,3*], Henrique G. Rosa[2*], Ivan Verzhbitskiy[2,3], Manuel J. L. F. Rodigues[2,4], Takashi Taniguchi[5], Kenji Watanabe[5], Goki Eda[2,3,4], Vitor M. Pereira[2,3,Φ], José Carlos Viana Gomes[2,3,◊]

[1]NUS Graduate School for Integrative Sciences and Engineering (NGS), University Hall, Tan Chin Tuan Wing, Level 4 #04-02, 21 Lower Kent Ridge, Singapore 119077

[2]Centre for Advanced 2D Materials (CA2DM), National University of Singapore, 6 Science Drive 2, Singapore 117546

[3]Department of Physics, National University of Singapore, 2 Science Drive 3, Singapore 117551

[4]Center of Physics and Department of Physics, Universidade do Minho, 4710-057, Braga, Portugal

[5]National Institute for Materials Science, 1-1 Namiki, Tsukuba 305-0044, Japan

[6]Department of Chemistry, National University of Singapore, 3 Science Drive 3, Singapore 117543

*These authors contributed equally to this work

[Φ]Corresponding author email: vpereira@nus.edu.sg

[◊]Corresponding author email: phyvjc@nus.edu.sg

# Contents

# 1 Sample fabrication and characterization

The sample used in this work is a mechanically exfoliated monolayer molybdenum disulfide ($MoSe_2$) flake transferred onto a fused silica substrate. In order to insulate the $MoSe_2$ sample from the environment, it was covered with a 20 nm thick flake of hexagonal-boron-nitride (hBN). Basic characterizations of the sample, such as optical microscopy, Raman spectroscopy and photoluminescence spectroscopy, are shown in Figure S1.

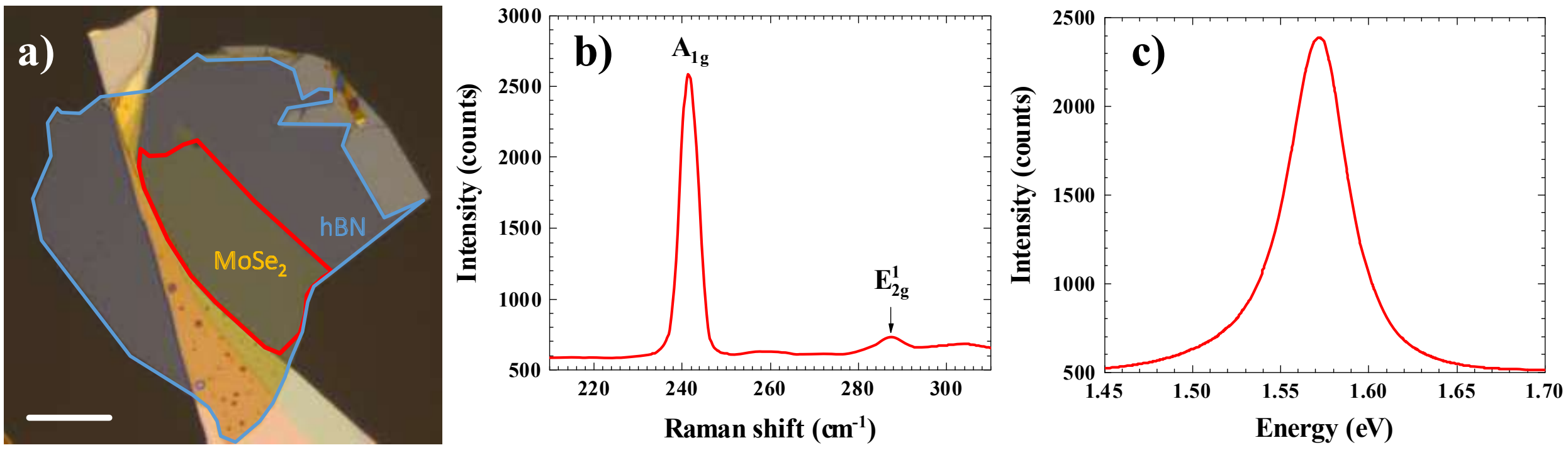

Figure S1: $MoSe_2$ flake characterization: a) Optical microscopy image (scale bar = 25 µm), b) Raman spectrum and c) photoluminescence spectrum (PL emission peak = 1.571 eV, FWHM = 40 meV).

From Figure S1c, it is possible to observe that the photon energy in the experiment, $E_{photon}$ = 1.590 eV, is slightly higher (19 meV) than the optical bandgap energy, shown by the PL emission peak at 1.571 eV.

## 2 Detailed data analysis

To obtain the actual helicity of the incoming light beam $\gamma$, the pump data are globally fitted to

$$I_{\text{pump}} = \frac{I_0}{2}[1 + \cos(2(\alpha_0 - \beta))\cos(2\gamma)]$$

with $I_0$ and $\alpha_0$ being the global variables. Here, $I_0$ is a normalizing factor for the intensities. Meanwhile, the SHG data (e.g. Figure 2b in main text) are fitted to

$$I_{\text{SH}} = I_0[\cos^2(\beta - \beta_{\text{max}}) + R\sin^2(\beta - \beta_{\text{max}})]$$

where $R = \frac{\text{minimum intensity}}{\text{maximum intensity}}$. Next, from our model (described below or in the main text) with the assumption of no VP, we can infer $\beta_{\text{max}}(\chi_{\text{vp}} = 0) = 0$. Therefore, a shift in SHG maximum angle $\Delta = \beta_{\text{max}}(\chi_{\text{vp}} \neq 0) - \beta_{\text{max}}(\chi_{\text{vp}} = 0) \neq 0$ indicates the presence of VP.

## 3 Conventional polarized second-harmonic generation

In order to identify the armchair (AC) directions of the $MoSe_2$ sample crystal, we performed conventional linear polarization resolved SHG measurements. In this case, the sample is placed between two parallel linear polarizers and the SHG power is measured as the polarizers are jointly rotated with respect to the sample. The AC directions correspond to the maxima of the resulting six-fold pattern.

As described in the main text, the data presented therein for the rotation of the SHG pattern was collected using a protocol (Protocol I) where the polarization state of the incoming laser beam is held fixed, and the linear analyzer is rotated by 360°. In this case the rotation is given by the angle $\Delta$ of Eq. (2) in the main text.

In addition, we have measured the SHG rotation using a second protocol (Protocol II), where the laser beam is linearly polarized at angle $\theta_1$ before reaching the quarter-wave plate (QWP, which sets the helicity $\gamma$ of the beam that reaches the sample), giving EP for the pump beam with $\alpha = \theta_1 + \gamma$ (see Figure 2a in the main text). The emitted SH signal is analyzed with a linear polarizer (P2), at angle $\theta_2$, such that $\alpha = \theta_2$. In this protocol, the rotation of the SHG pattern is given by the angle $\Delta'$ described in the main text under "Linearly polarized SHG".

Figure S2 shows that we obtain the same trend with increasing $\gamma$. Importantly, Figure S2b shows that the measured values of $\Delta$ (obtained from Protocol I) are approximately 3 times those of $\Delta'$ (Protocol II), which is precisely what we anticipate from the discussion preceding Eq. (2) in the main text.

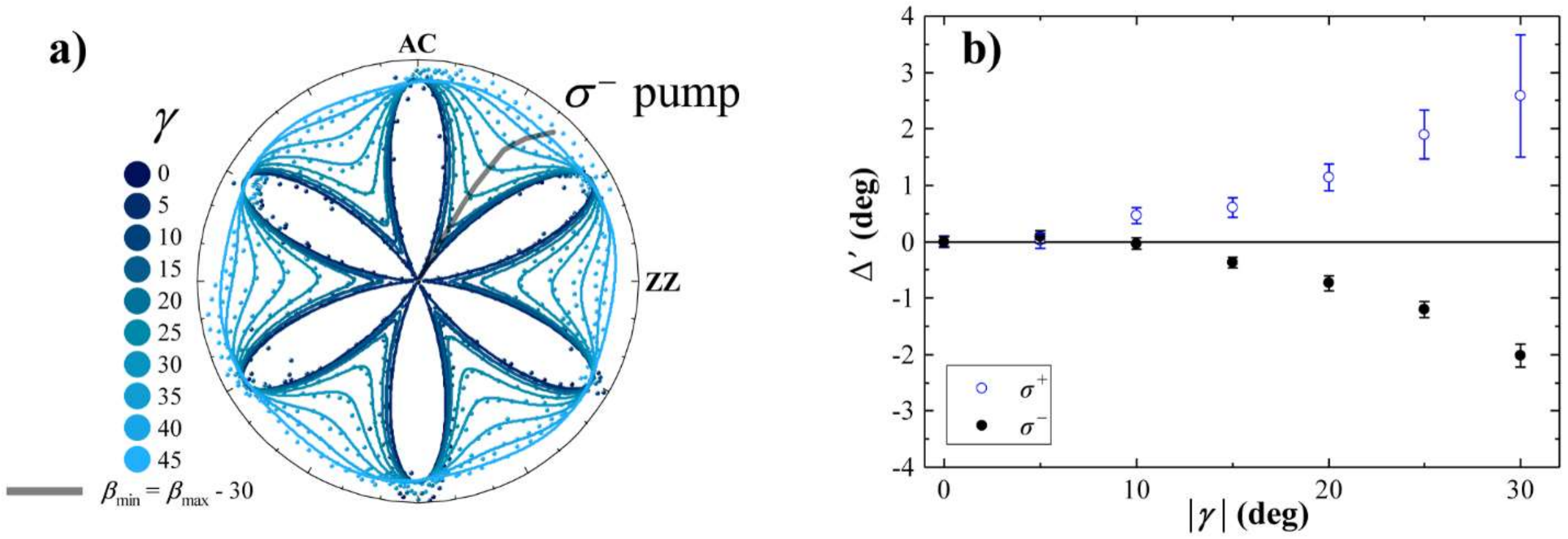

Figure S2: Conventional pSHG scheme. a) 6-fold symmetry of pSHG of $MoSe_2$ for $\gamma$ in the range of 0 to 30°. Grey line as a guide for the eye for the minimum SHG angle across $\gamma$. b) Maximum SHG angle offset compared to $\gamma = 0$ against $\gamma$ for both $\sigma^+$ (circles) and $\sigma^-$ (disks) pump excitation. We use $\Delta'$ for the SHG pattern rotation angle obtained with Protocol II to distinguish it from $\Delta = 3\Delta'$ used in the main text (Protocol I).

# 4 Nonlinear optical susceptibilities

The $\chi_{\text{int}}$ and $\chi_{\text{vp}}$ are the intrinsic and valley polarization second-order susceptibilities, defined as:

$$\chi_{\text{int}} \equiv \chi_{111} = -\chi_{122} = -\chi_{212} = -\chi_{221}$$
$$\chi_{\text{vp}} \equiv \chi_{222} = -\chi_{211} = -\chi_{121} = -\chi_{112}$$

where the Cartesian directions 1 and 2 are along the crystal's zig-zag and armchair directions, respectively.

## 5 SH offset angle model

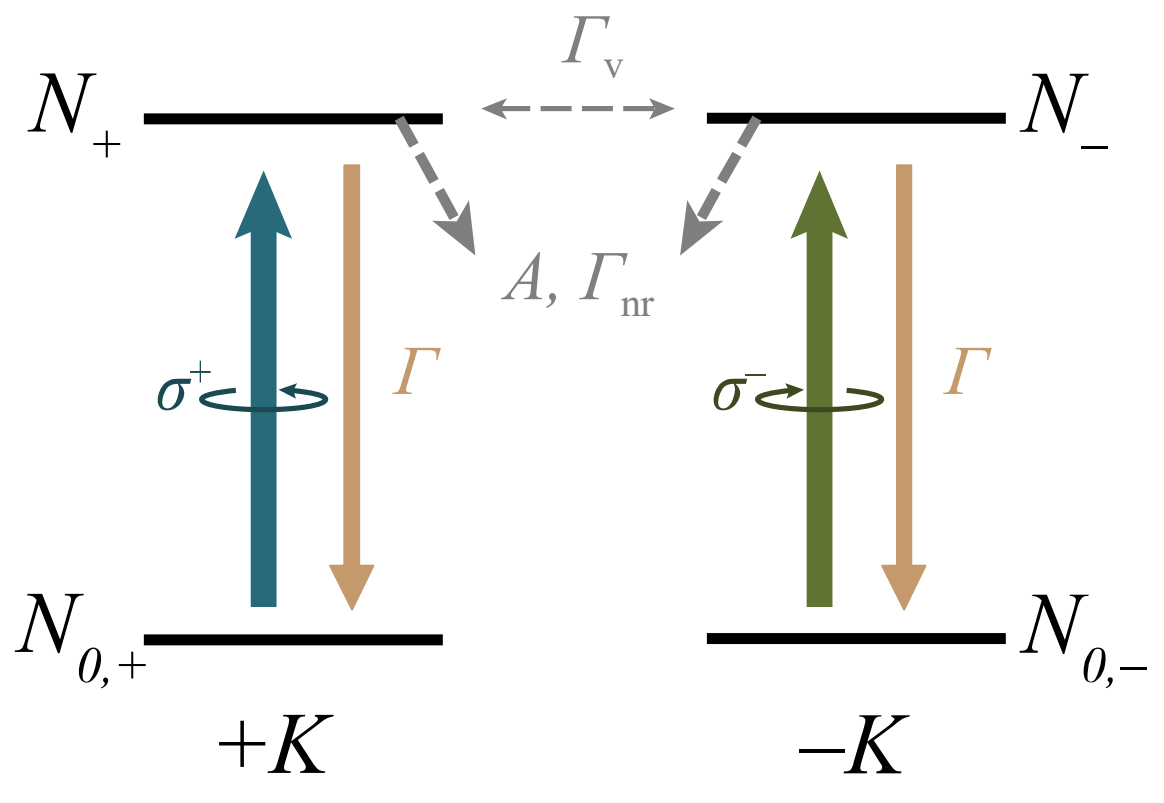

Figure S3: A pair of 2-level systems to model the transient population dynamics at the ±K valleys.

To model the exciton population dynamics, we consider a pair of 2-level systems (Figure S3) with 2 degenerated excited states representing $\pm K$ valleys ($N_+$ and $N_-$) and their ground states ($N_{0,+}$ and $N_{0,-}$). Taking the recombination rates of the valleys to be the same ($\Gamma_+ = \Gamma_- = \Gamma$) and a second-order term ($AN_\pm^2$) to account for two-body annihilation, we can write the rate equations as

$$\frac{\mathrm{d}N_\pm}{\mathrm{d}t} = -(\Gamma + \Gamma_{\mathrm{nr}})N_\pm - AN_\pm^2 - B\frac{S(\omega)}{c}I_\pm\left(N_+ - N_{0,+}\right) + \Gamma_v(N_\mp - N_\pm) \tag{1}$$

where $N_\pm$, $N_{0,\pm}$ and $N_T$ are the populations in the excited, ground states of $\pm K$ valley and total population in each valley, respectively. $\Gamma_{\mathrm{nr}}$ is the nonradiative decay rate. $B$, $S(\omega)$ and $c$ are the Einstein $B$ coefficient, transition linewidth and speed of light in vacuum. Using the relationship between $\Gamma$ and $B$

$$B = \Gamma\frac{c^3}{8\pi h\nu^3 n^3} \tag{2}$$

where $n$ is the refractive index, and introducing the absorption cross section $\sigma(\omega)$,

$$\sigma(\omega) = \left(\frac{\lambda}{n}\right)^2\frac{\Gamma\, S(\omega)}{8\pi}, \tag{3}$$

the rate equations can be rewritten in the form

$$\frac{\mathrm{d}N_\pm}{\mathrm{d}t} = -(\Gamma + \Gamma_{\mathrm{nr}})N_\pm - AN_\pm^2 - C\, I_\pm(N_T - 2N_\pm) + \Gamma_v(N_\mp - N_\pm) \tag{4}$$

where $C = \frac{\sigma(\omega)}{n\,\hbar\omega}$ together with $N_T = N_\pm + N_{0,\pm}$ where $N_T$ is the total density of available electron-hole excitations given the parameters of the laser excitation, which is estimated from the density of state ($D$) and laser linewidth ($N_T = D\,\delta\hbar\omega$). Besides, the difference in intensities of $\sigma^+$ and $\sigma^-$ ($I_+$ and $I_-$ respectively) can be expressed in terms of the degree of ellipticity, $\gamma$ (see main text for definition) as $\frac{I_+-I_-}{I_++I_-} = \frac{I_+-I_-}{I} = \sin(2\gamma)$, or

$$I_\pm = I\frac{1 \pm \sin 2\gamma}{2}. \tag{5}$$

Solving these equations will provide the population imbalance ($\Delta N = N_+ - N_-$) in the steady state or transient limits, as shown in the next section.

Next, to connect $\chi_{\rm vp}$ to $\Delta N$, we assume

$$\frac{\chi_{\rm vp}}{\chi_{\rm int}} = k_{\chi\mu}\Delta\mu \tag{6}$$

where $k_{\chi\mu}$ is a proportionality constant and $\Delta\mu$ is the Fermi level difference between the valleys [1,2]. To connect $\Delta N$ to $\Delta\mu$, we first integrate $N = \int_0^\infty D\, f(E)\, dE$, where $D = \frac{m^*}{\pi\hbar^2}$ and $f(E)$ are the density of states of $MoSe_2$ and Fermi-Dirac distribution, which gives $N = Dk_BT\log(1 + e^{\frac{\mu}{kT}})$. The relationship is obtained from $\Delta\mu \approx \frac{\partial\mu}{\partial N}\Delta N$, giving

$$\Delta\mu \approx \frac{k_B}{D}\frac{1}{1 - e^{-\frac{N}{D\,T}}}\Delta N \equiv k_{\mu N}(T)\,\Delta N. \tag{7}$$

We noted that $k_{\mu N}$ increases slowly with temperature, giving us larger sensitivity at higher temperature. Finally, by modelling our experimental setup and procedure described in the Experimental Setup, the SHG signal as a function the analyze angle $\beta$, measured with respect to the major axis of the incoming polarization ellipse is

$$I_{SHG} \propto \left[\sin^2 2\gamma\left(\chi_{\rm int}\sin\beta + \chi_{\rm vp}\cos\beta\right)^2 + \left(\chi_{\rm int}\cos\beta - \chi_{\rm vp}\sin\beta\right)^2\right] I^2 \tag{8}$$

where $\chi_{\rm vp}$ and $I$ will be functions of time in the transient description that we elaborate below. Also, according to the discussion in the main text, we consider $\chi_{\rm int}$ and $\chi_{\rm vp}$ as real quantities. The rotation of the SHG pattern (the angle $\Delta$) in the presence of VP is obtained from this result by solving $\frac{dI_{SHG}}{d\beta} = 0$ to give

$$\tan\Delta = \frac{\chi_{\rm vp}}{\chi_{\rm int}}. \tag{9}$$

## 5.1 Steady state solution

Setting the time derivatives in the rate equations to zero, we can straightforwardly calculate the steady state populations in the presence of a constant irradiance at the sample $I$, $A = 0$ and $\Gamma_v = 0$ (which is what we are using, explained below). We obtain

$$\Delta N_{SS} = N_+ - N_- = N_T \frac{C\, I\, (\Gamma + \Gamma_{\mathrm{nr}})\, \sin(2\gamma)}{(C\, I + \Gamma + \Gamma_{\mathrm{nr}})^2 - [C\, I \sin(2\gamma)]^2} \tag{10}$$

with $C = \frac{\sigma(\omega)}{n\, \hbar\omega} = \frac{1 - \ln a}{n\, \hbar\omega\, N_T}$ where $a = 6\%$ is the absorbance at low intensity [3]. To visualize the behavior of $\Delta N_{SS}$ as a function of $\gamma$, setting intensities to be the average and peak intensities used in the experiment, the $\Delta N_{SS}$ are shown in Figure S4a and b, respectively. The other parameters are $\Gamma = 1/0.3061\ \mathrm{ps}^{-1}$, $\Gamma_{\mathrm{nr}} = 1/0.0383\ \mathrm{ps}^{-1}$ at room temperature [4], $N_T = \frac{0.25\, m_e}{\pi\hbar} \delta\omega_{\mathrm{laser}} = 2.1 \times 10^{12}$ cm$^{-2}$ for laser linewidth of $\delta\omega_{\mathrm{laser}} = 20$ meV, $n = (2.24 + 1.45)/2$ is the mean refractive index between hBN and $SiO_2$ and $\hbar\omega = 1.6$ eV. Therefore, $C = 3.86 \times 10^3$ cm J$^{-1}$ in Equation (10) for $I$ and $\Gamma_{(\mathrm{nr})}$ in units of GW cm$^{-2}$ and ps$^{-1}$, respectively.

In the regime of small $I$ (i.e. $I \ll \frac{\Gamma + \Gamma_{\mathrm{nr}}}{C} \equiv I_m = 7.6$ MW cm$^{-2}$), $\Delta N_{SS} \approx N_T \frac{C\, I}{\Gamma + \Gamma_{\mathrm{nr}}} \sin 2\gamma \propto I$. In the other extreme, the large $I$ regime gives $\Delta N_{SS} \approx N_T \frac{\Gamma + \Gamma_{\mathrm{nr}}}{C\, I} \tan 2\gamma \sec 2\gamma \propto I^{-1}$.

This steady state calculation fails to capture the trend of the experimental data presented in the main text: From Figure S4a, which is calculated using the mean intensity of the pulse laser, we can see that the steady state result predicts that the population imbalance should grow at a slower rate for large $\gamma$; On the other hand, in Figure S4b, calculating with peak intensity gives a trend similar to the experimental data, but the calculation shows that the population imblance should increase with the incident intensity, which is not the case.

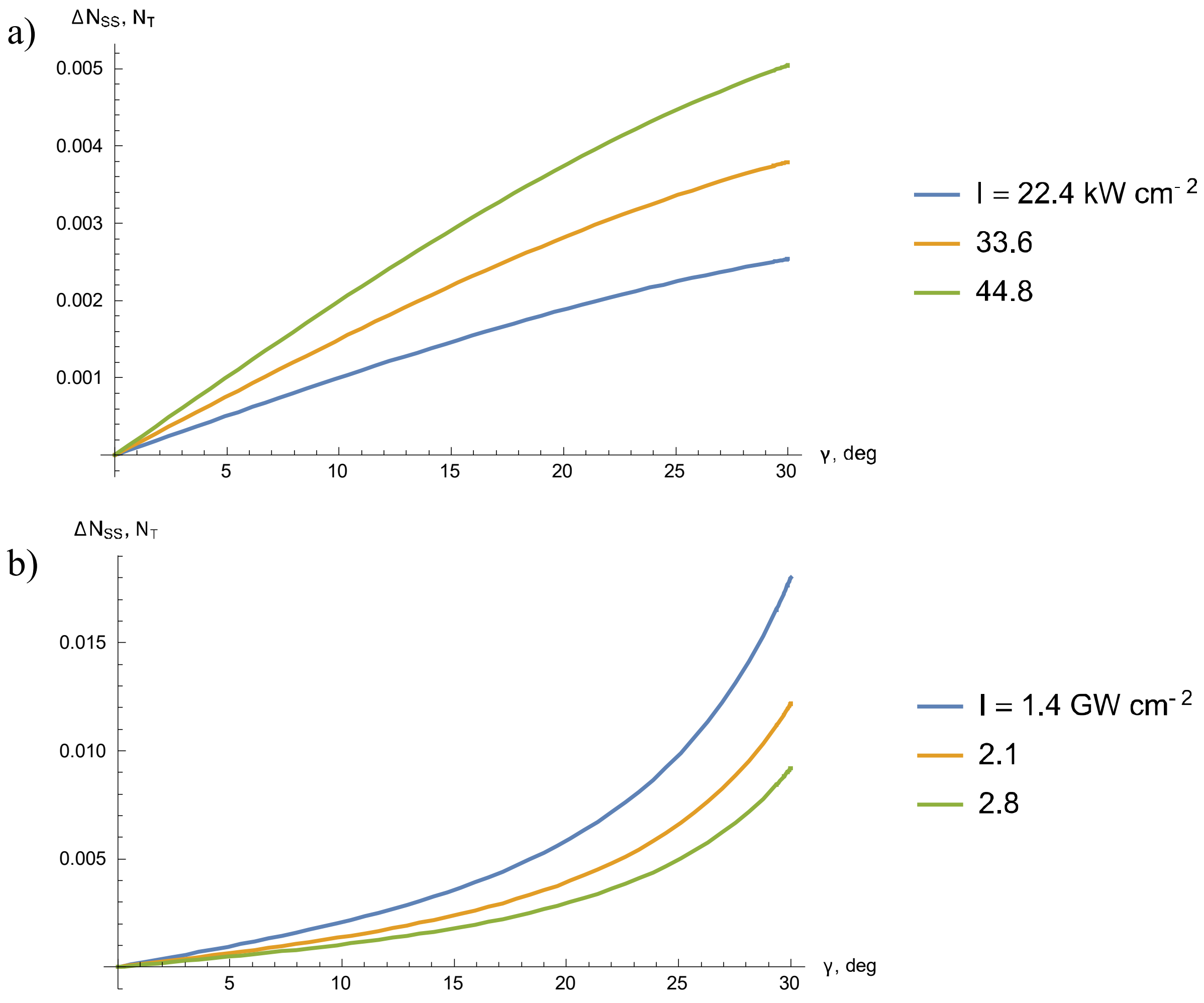

Figure S4: Behavior of the steady-state solution, $\Delta N_{SS}$, as given in Equation (10). The parameter $I$ has been set to the mean (a) and peak (b) intensities used in the experiment.

### 5.1.1 Steady state solution without exciton-exciton annihilation

For reference, we solved the rate equation with no exciton-exciton annihilation term ($A = 0$) because of its simplicity, and also because the full transient behavior of the population imbalance was found to be described by the same functional form of the steady state solution with $A = 0$ (to be discussed in the following section, below). The overall behavior of $\Delta N_{SS,A\neq 0}$ is the same as $\Delta N_{SS}$ but scaled by a negligible factor in the appropriate range, as illustrated in Figure S5.

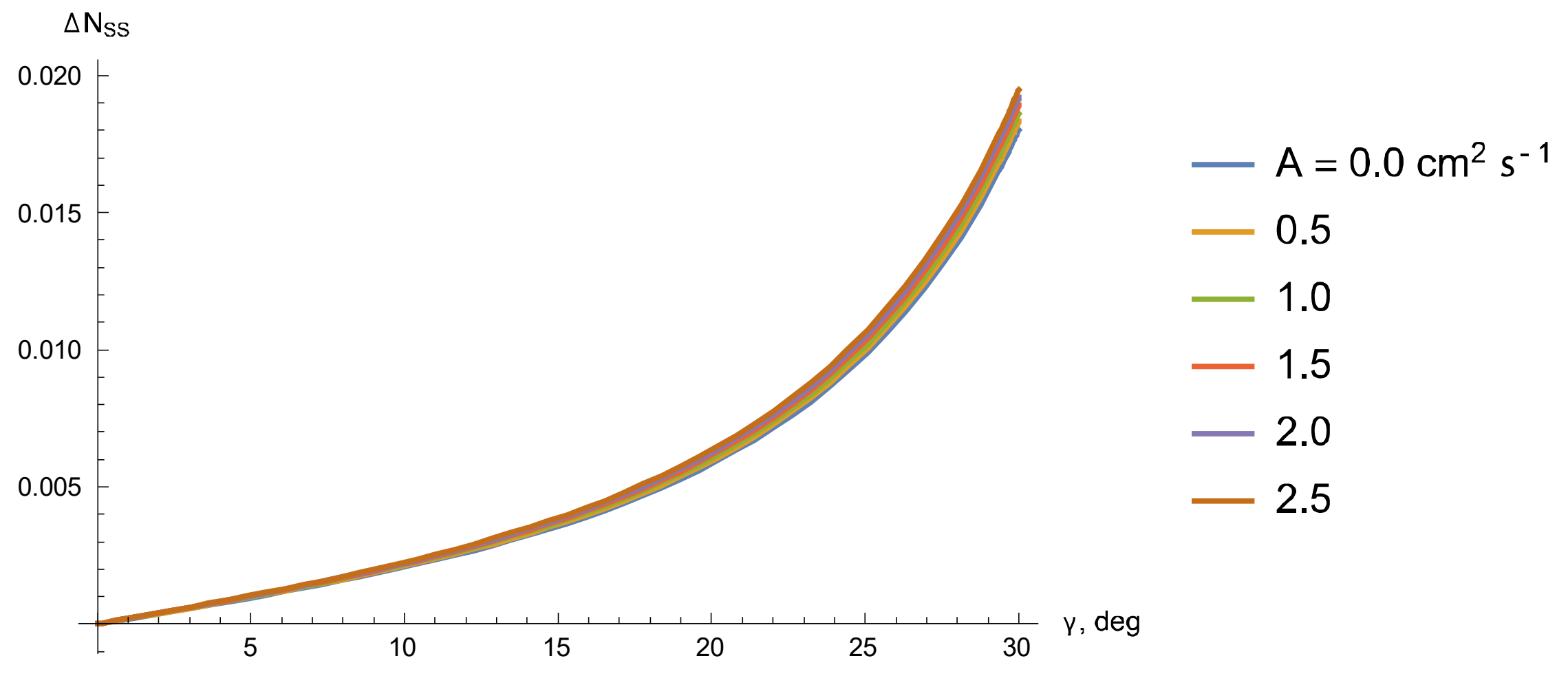

Figure S5: $\Delta N_{SS}$ with different exciton-exciton annihilation rates for $I = 1.4\ \mathrm{GW\ cm^{-2}}$.

## 5.2 Transient solution

The steady state solution does not fully model our experimental conditions, because we used an ultrafast pulsed laser and much higher peak intensities in comparison with the CW saturation intensity. Therefore, we solved Equation (4) to obtain the full time dependence of the exciton populations in response to a single pulse impinging the sample (considering the response to an isolated pulse is entirely justified by the fact that successive pulses are separated by 1/(80 MHz) = 12.5 ns, while all relaxation processes happen in the order of, at most, a few ps). For this numerical calculation, unless otherwise stated, we used $\Gamma = 1/0.3061\ \mathrm{ps}^{-1}$, $\Gamma_{\mathrm{nr}} = 1/0.0383\ \mathrm{ps}^{-1}$ [4], $A = 0.33 \times 10^{12}\ \mathrm{cm}^2\ \mathrm{ps}^{-1}$ [5], $N_T = 2.1 \times 10^{12}\ \mathrm{cm}^{-2}$, $n = 1.845$, $\hbar\omega = 1.59$ eV and $\Gamma_v = 0$; the latter assumption equates to neglecting the intervalley exciton relaxation, which is justified by the fact that our photon energy is only 19 meV above the center of the PL peak, which is less than the threshold energy (38 meV for $MoSe_2$) for phonon-assisted intervalley scattering [6]. Figure S6 shows the calculation results for low, optimum and high intensities. The laser pulse is modelled as $I = I_0 f(t)$ with $f(t) = e^{-(t/\tau)^2}$ and $\tau$ being the pulse duration.

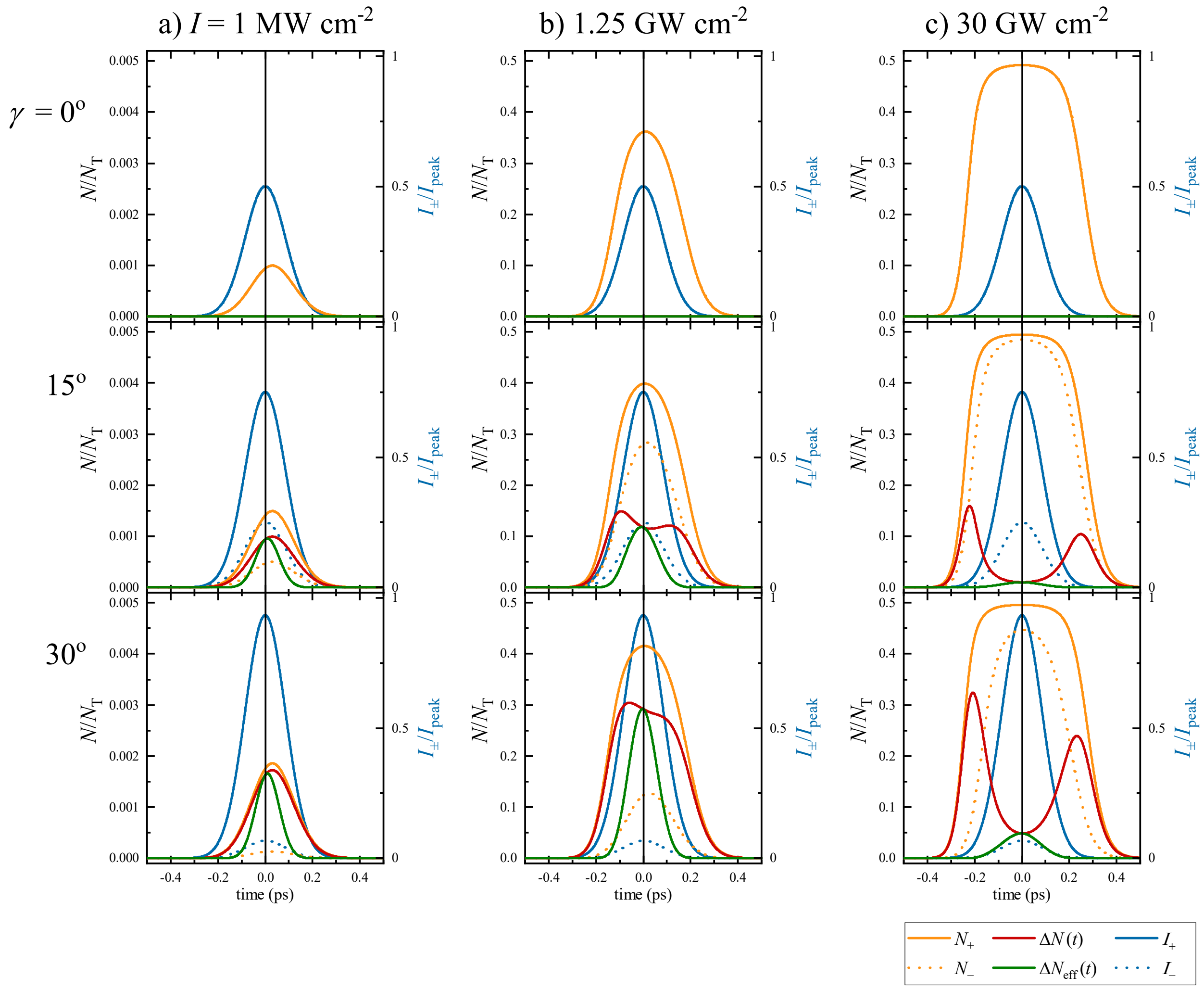

Figure S6: Calculation of populations in $\pm K$ valleys (orange solid and dashed) for $I_0$ = a) 1 MW cm$^{-2}$ (low), b) 1.25 GW cm$^{-2}$ (optimum) and c) 30 GW cm$^{-2}$ (high intensity) for $\gamma = 0°, 15°$ and $30°$. $I_\pm$ is the normalized $\sigma^\pm$ intensity, $N_\pm$ is the population in $\pm$K valley, $\Delta N(t) = N_+ - N_-$ is the transient population imbalance and $\Delta N_{\text{eff}}(t) = \Delta N(t) I(t)$ is the $\Delta N(t)$ normalized with laser pulse.

As the detector collecting the SHG performs a temporal integration of the transient SHG emission, the rotation ($\Delta$) of the SHG pattern that we measured reflects not the instantaneous (transient) SHG intensity, but a value averaged over time. To capture this, and recalling that the SHG intensity is proportional to the square of the incident beam's intensity, Equation (9) is generalized to a time-average weighted by the temporal profile of the incident pulse:

$$\tan \Delta = \frac{\int \left(\chi_{\text{vp}}(t)\right) \, f^2(t) \, dt}{\int (\chi_{\text{int}}) \, f^2(t) \, dt}. \tag{11}$$

Using the relation

$$\frac{\chi_{\mathrm{vp}}(t)}{\chi_{\mathrm{int}}} = \kappa\,\Delta N(t), \tag{12}$$

Equation (11) can be simplified as

$$\tan\Delta = \kappa\frac{\int \Delta N(t)\, f^2(t)\, dt}{\int f^2(t)\, dt} \equiv \kappa\,\Delta N_{\mathrm{eff}}. \tag{13}$$

This shows that the rotation of the SHG pattern due to the valley imbalance can be expressed in terms of an effective population imbalance, $\Delta N_{\mathrm{eff}}$, which corresponds to the temporal average of the transient imbalance weighted by the square of the pulse's lineshape. This is presented as Equation (4) in the main text.

The analysis of how different excitation conditions impact the rotation angle $\Delta$ can thus be made by studying the conceptually simple quantity $\Delta N_{\mathrm{eff}}$. We now describe the behavior of $\Delta N_{\mathrm{eff}}$, computed from the numerical integration of Equation (4), and how the results allow us to extract the proportionality constant $\kappa$ by fitting the experimental data for $\Delta$ as a function of $\gamma$.

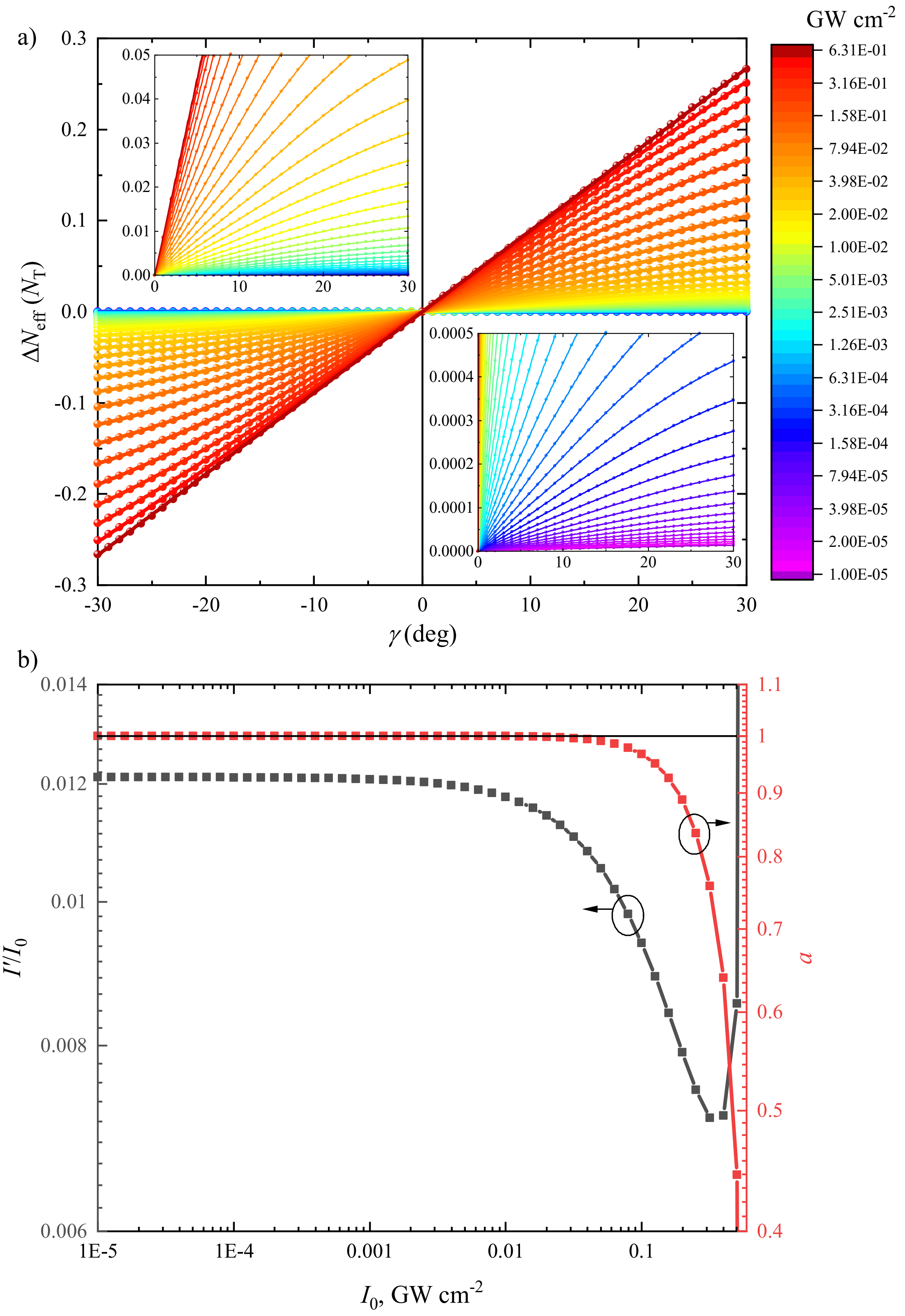

Figure S7: a) Calculation summary of the population imbalance as a function of ellipticity for low pump intensities. The insets provide a close-up of the main plot on a smaller vertical range. The points convey $\Delta N_{\text{eff}}$ obtained from the numerical integration of Equation (4). The lines are fits to $\frac{\Delta N_{\text{eff}}}{N_T} = I'/I_m \sin 2a\gamma$. b) Ratio of fitted $I_0$ to $I_0$ ($I'/I_0$) and fitted $a$ for various $I_0$ used in the calculation.

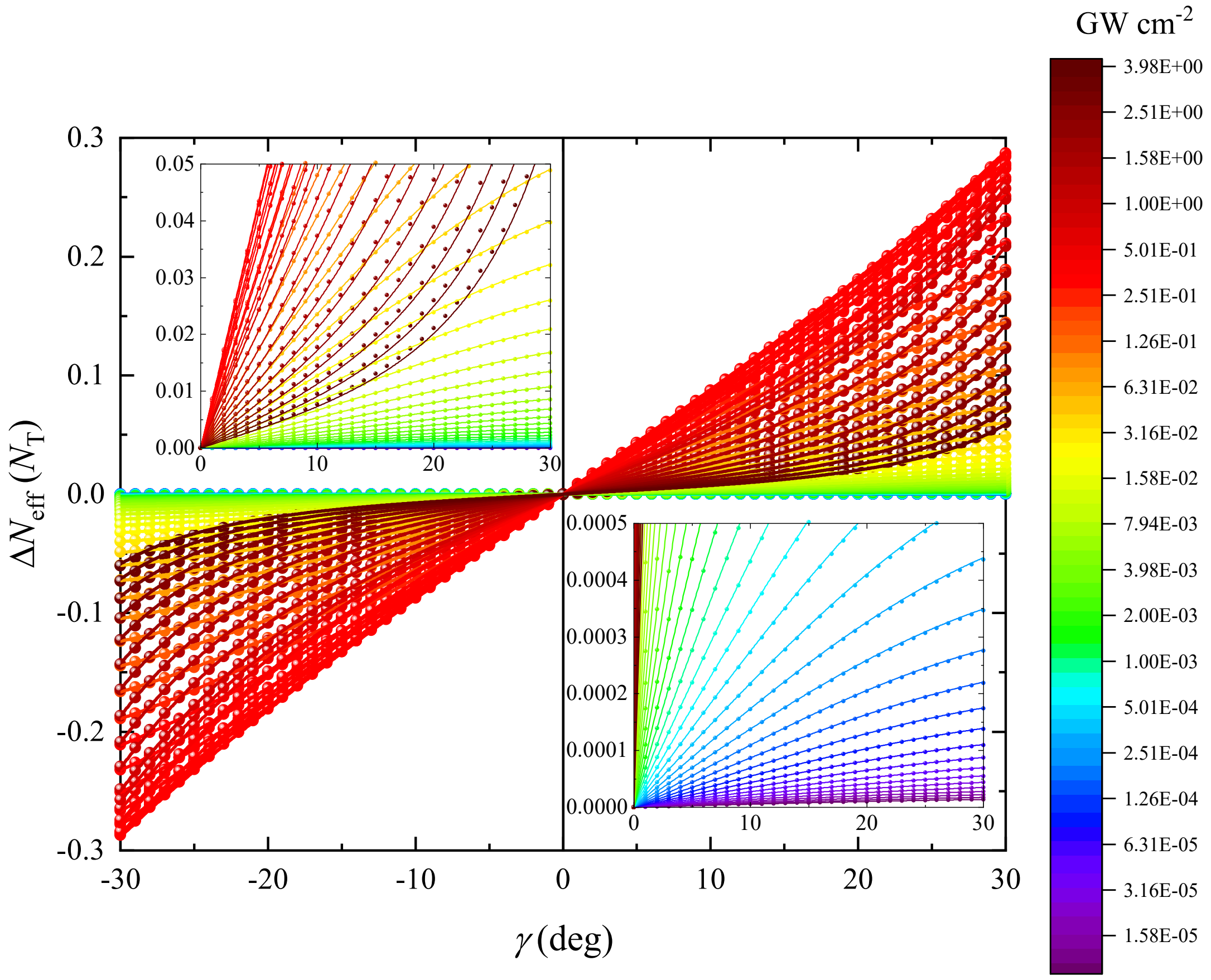

Figure S8: Calculation summary of the population imbalance as a function of ellipticity for different $I_0$. The points convey $\Delta N_{\mathrm{eff}}$ obtained from the numerical integration of Equation (4). The entire dataset (for all $\gamma$ and all intensities $I$) was fit to Equation (14) in order to obtain the global fitting parameter $b = 0.012217 \pm 3 \times 10^{-6}$. Each line corresponds to $\Delta N_{\mathrm{eff}}$ as given in Equation (14) for each intensity.

Figure S7 and Figure S8 show the calculation summaries of $\Delta N_{\mathrm{eff}}$ as a function of $\gamma$ for low and high $I_0$, respectively. From Figure S7a, we can see that it correctly captures to the very low intensity approximation of $\Delta N_{\mathrm{eff}} \propto \sin 2\gamma$, which is just the difference in $\sigma^{\pm}$ intensities; when $I_0 \gtrsim 0.01$ GW cm$^{-2}$, this simplification starts to fail because, as we can see in Figure S7b, the fitting parameter $a$ (in $\sin 2a\gamma$) and $I'/I_0$ start to deviate from unity and from the small $I_0$ value, respectively.

We find, however, that the results in Figure S8 (obtained from the full time-dependent solution of the rate equations) can be very well described by the following expression

$$\frac{\Delta N_{\mathrm{eff}}}{N_T} = \frac{b\frac{I_0}{I_m}\sin(2\gamma)}{\left(b\frac{I_0}{I_m}+1\right)^2 - \left[b\frac{I_0}{I_m}\sin(2\gamma)\right]^2} \tag{14}$$

with fixed $I_m = \frac{\Gamma+\Gamma_{\mathrm{nr}}}{c} = 7.6$ MW cm$^{-2}$. This corresponds to the steady state solution presented earlier in Equation (10), except that the peak intensity, $I_0$, appears now renormalized by the parameter $b$. All the curves Figure S8 can be fitted to this expression using a single (global) value of the parameter $b$, namely $b = 0.012217 \pm 3 \times 10^{-6}$. This can be interpreted as a reduction by a factor of ~100 in the effective intensity that contributes to the population imbalance, in comparison with a CW excitation with intensity $I_0$.

Figure S9 shows the response of the population imbalance ($\Delta N_{\mathrm{eff}}$ at $\gamma = 30°$) with $I_0$ at different temperature (the effect of temperature is captured by varying the nonradiative lifetime [4]), with the values of $I_0$ used in the experiment highlighted in blue. At room temperature, the pump intensity used in the experiment gives the largest valley population imbalance, as illustrated by the fact that the maximum of the red curve in Figure S9 occurs within the experimental window of intensities. The total population imbalance increases as we reduce the temperature, but one should note that the optimum intensity $I_{\mathrm{opt}}$ to reach maximum population imbalance is progressively reduced with decreasing temperature (from $I_{\mathrm{opt}}$ ~ 1.3 GW cm$^{-2}$ at 300 K to $I_{\mathrm{opt}}$ ~ 0.3 GW cm$^{-2}$ at 4 K).

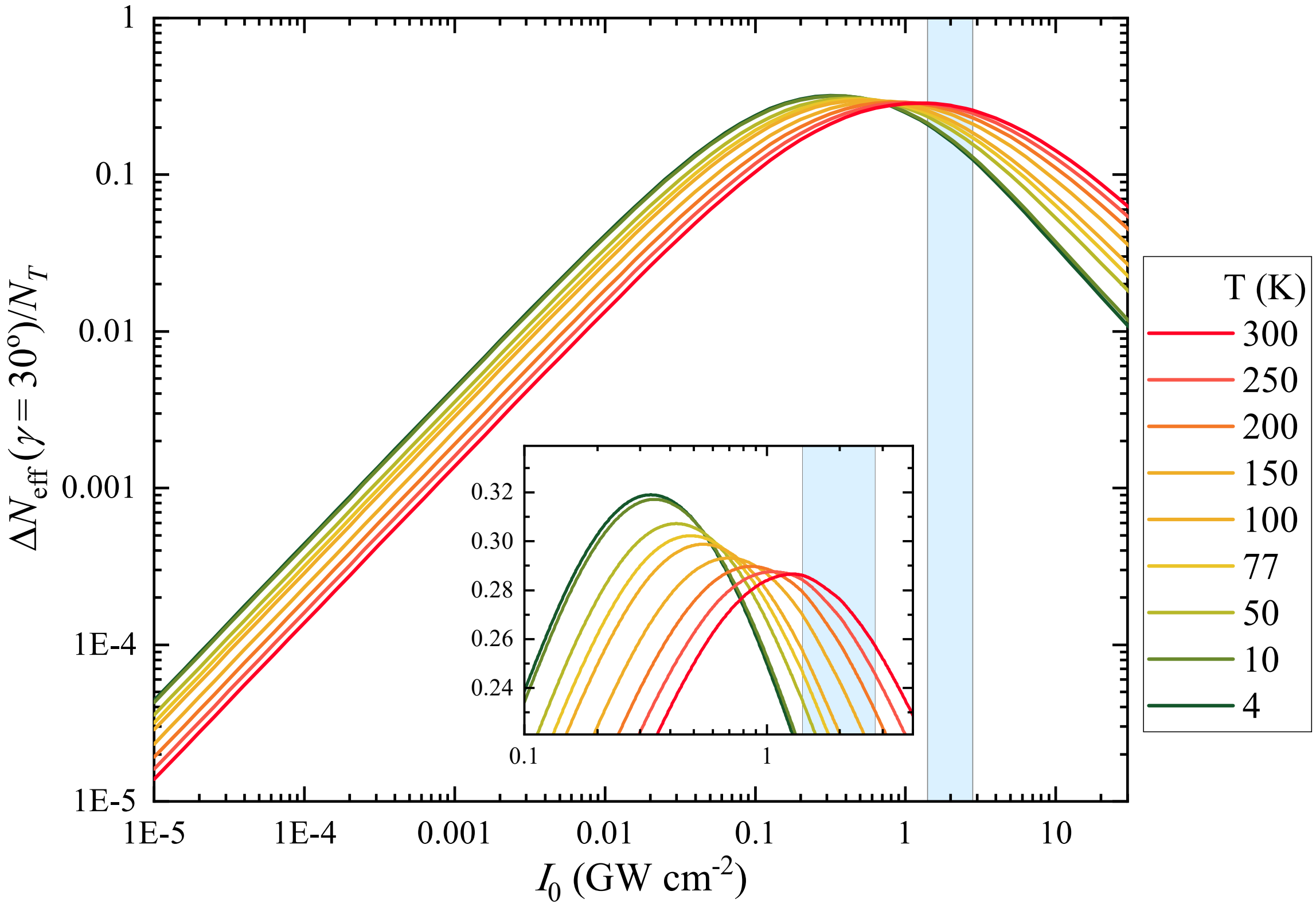

Figure S9: Calculated population imbalance as a function of laser peak intensity ($I_0$), at different temperatures, and for helicity $\gamma = 30°$. The range of intensities used in the experiment is highlighted by the blue area. The inset magnifies the region near the maximum of the curves in the main panel.

Figure S10 shows the time-averaged population difference between the ground and the excited states of a two-level system (i.e. by setting $\gamma = 45°$ in Equation (4) and solving only for $N_{+}$), considering three different scenarios: (i) transient and without normalization (black), (ii) transient and normalized with laser intensity profile (orange) and (iii) steady state (blue), which are proportional to the photon absorption, as a function of pump intensity. In steady state, $I_{\text{sat}}$ is defined as the intensity at which the absorption decreases to half of the low intensity absorption (i.e. $N_0 - N_1 = 0.5\ N_T$) and corresponds to the expected saturation intensity under CW excitation. However, when we consider the full transient solution, the actual value of $I_{\text{sat}}$ increases to a higher value (using the same parameters as in Figure S6 and those used in the main text), delaying the onset of population saturation (and therefore VP) being probed. This explains our observation of VP in the experiments, despite having the incoming beam with peak intensities up to an order of magnitude higher than $I_{\text{sat}}$.

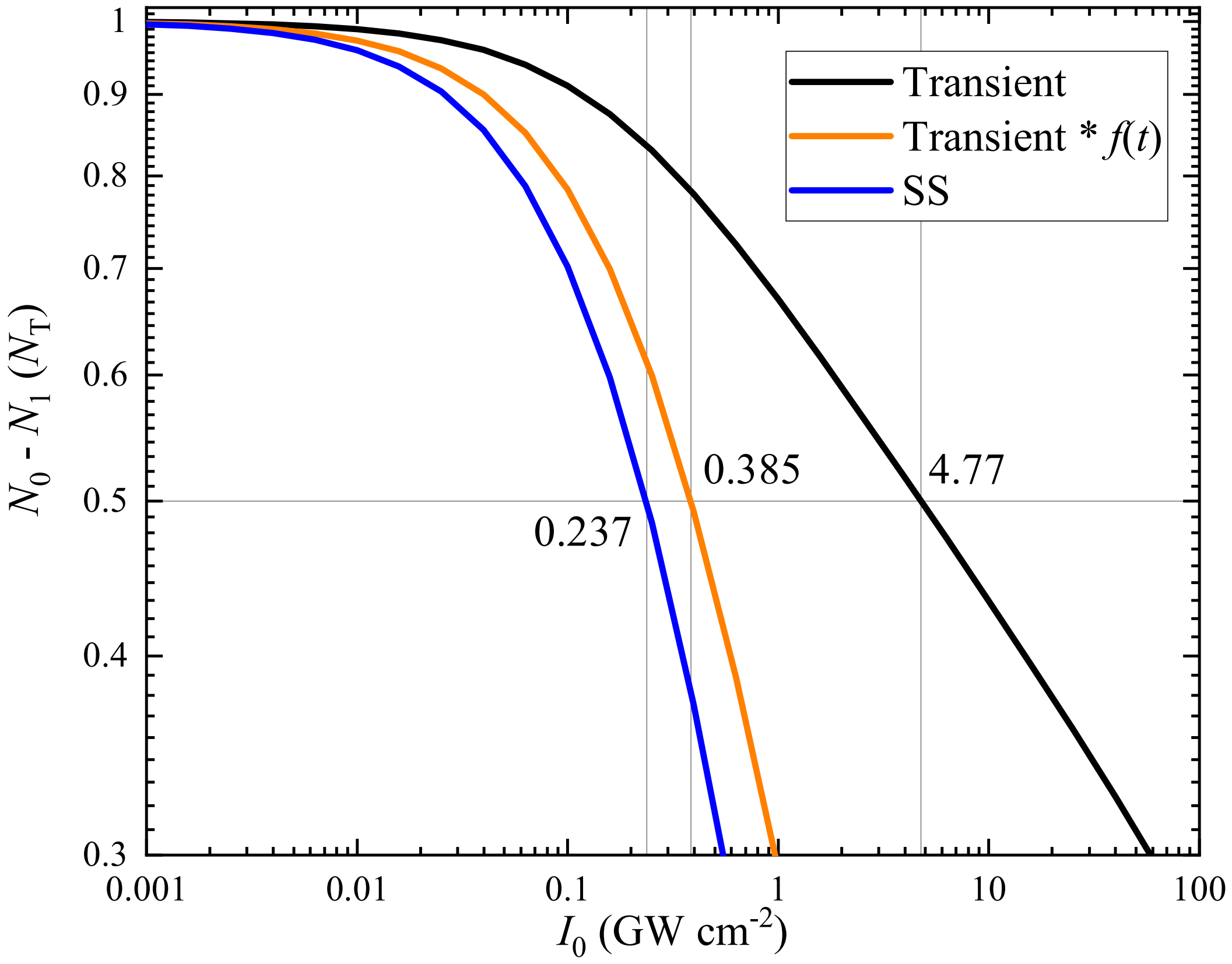

Figure S10: Time-averaged difference between the ground and excited states population ($N_0 - N_1$) without normalization (black), normalized with laser intensity profile (orange) and the steady state value (blue). They correspond to the absorption as a function of pump intensity. We obtain $I_{\text{sat}} = 0.237$, 0.385 and 4.77 GW cm$^{-2}$ for the 3 different scenarios described in the text.

## 6 Supplementary references

1. L. E. Golub and S. A. Tarasenko, "Valley polarization induced second harmonic generation in graphene," Phys. Rev. B **90**, 201402 (2014).

2. T. O. Wehling, A. Huber, A. I. Lichtenstein, and M. I. Katsnelson, "Probing of valley polarization in graphene via optical second-harmonic generation," Phys. Rev. B **91**, 041404 (2015).

3. Z. Nie, C. Trovatello, E. A. A. Pogna, S. Dal Conte, P. B. Miranda, E. Kelleher, C. Zhu, I. C. E. Turcu, Y. Xu, K. Liu, G. Cerullo, and F. Wang, "Broadband nonlinear optical response of monolayer $MoSe_2$ under ultrafast excitation," Appl. Phys. Lett. **112**, 031108 (2018).

4. M. Selig, G. Berghäuser, A. Raja, P. Nagler, C. Schüller, T. F. Heinz, T. Korn, A. Chernikov, E. Malic, and A. Knorr, "Excitonic linewidth and coherence lifetime in monolayer transition metal dichalcogenides," Nat. Commun. **7**, (2016).

5. N. Kumar, Q. Cui, F. Ceballos, D. He, Y. Wang, and H. Zhao, "Exciton-exciton annihilation in $MoSe_2$ monolayers," Phys. Rev. B - Condens. Matter Mater. Phys. **89**, 1–6 (2014).

6. G. Kioseoglou, A. T. Hanbicki, M. Currie, A. L. Friedman, and B. T. Jonker, "Optical polarization and intervalley scattering in single layers of $MoS_2$ and $MoSe_2$," Sci. Rep. **6**, 25041 (2016).